\documentclass[12pt]{article}
\usepackage{bbm}
\usepackage{bbding}
\usepackage{amsmath}
\usepackage{amssymb}
\usepackage{amsfonts}
\usepackage{mathrsfs}
\usepackage{mathrsfs, amssymb}
\usepackage{booktabs}
\usepackage[perpage,symbol*]{footmisc}
\usepackage{graphicx}
\usepackage{multirow}
\usepackage[level]{datetime}
\usepackage[a4paper]{geometry}

\newcounter{theorem}
\newtheorem{theorem}{Theorem}[section]
\newcounter{lemma}

\newcounter{remark}

\newcounter{example}

\newcounter{definition}

\newcounter{corollary}

\newcounter{proposition}

\newcounter{assumption}

\newcounter{condition}

\newcounter{algorithm}

\makeatletter
\newcommand*{\indep}{%
  \mathbin{%
    \mathpalette{\@indep}{}%
  }%
}
\newcommand*{\nindep}{%
  \mathbin{
    \mathpalette{\@indep}{\not}
  }%
}
\newcommand*{\@indep}[2]{%
  \sbox0{$#1\perp\m@th$}
  \sbox2{$#1=$}
  \sbox4{$#1\vcenter{}$}
  \rlap{\copy0}
  \dimen@=\dimexpr\ht2-\ht4-.2pt\relax
  \kern\dimen@
  {#2}%
  \kern\dimen@
  \copy0 
}
\makeatother

\def\0{\mbox{\boldmath$0$}}
\def\H_N{\mbox{\boldmath$H_N$}}

\newcommand{\bmsection}[1]{%
	\section{\textbf{#1}}%
}

\begin{document}
\baselineskip 18pt
\begin{center}
{\large\bf Optional subsampling for generalized estimating equations in growing-dimensional longitudinal Data}
\end{center}

\begin{center}

{ \textbf{\small{ Chunjing Li}}}\\
{ \small{School of Mathematics and Statistics, Changchun University of Technology, China}}\\
{ \small{\texttt{\textsc{lichunjing@ccut.edu.cn}}}}

{ \textbf{\small{ Jiahui Zhang}}}\\
{ \small{School of Mathematics and Statistics, Changchun University of Technology, China}}\\
{ \small{\texttt{\textsc{19842836529@163.com}}}}

{ \textbf{\small{ Xiaohui Yuan}}$^*$}\\
{ \small{School of Mathematics and Statistics, Changchun University of Technology, China}}\\
{ \small{\texttt{\textsc{yuanxh@ccut.edu.cn}}}}

\end{center}

\begin{center}
This version: \usdate\today
\end{center}

\footnotetext{$^*$Corresponding author, $^\dag$ equal authors contribution.}

\begin{abstract}
As a powerful tool for longitudinal data analysis, the generalized estimating equations have been widely studied in the academic community. However, in large-scale settings, this approach faces pronounced computational and storage challenges. In this paper, we propose an optimal Poisson subsampling algorithm for generalized estimating equations in large-scale longitudinal data with diverging covariate dimension, and establish the asymptotic properties of the resulting estimator. We further derive the optimal Poisson subsampling probability based on A- and L-optimality criteria. An approximate optimal Poisson subsampling algorithm is proposed, which adopts a two-step procedure to construct these probabilities. Simulation studies are conducted to evaluate the performance of the proposed method under three different working correlation matrices. The results show that the method remains effective even when the working correlation matrices are misspecified. Finally, we apply the proposed method to the CHFS dataset to illustrate its empirical performance.
\end{abstract}

\section {Introduction} \label{sec1} \setcounter {equation}{0}
\def\theequation{\thesection.\arabic{equation}}

Longitudinal data are commonly encountered in medical research, economics, and the social sciences, and have garnered significant attention in statistical research. Liang \& Zeger (1986) developed generalized estimating equations (GEE) for the analysis of longitudinal data, extending quasi-likelihood approaches by incorporating a working correlation matrix to account for within-subject dependence. The resulting estimators remain consistent despite potential misspecification of the working correlation matrix. Chaganty (1997) demonstrated that the parameter estimates are consistent and asymptotically normal. Li (1997) studied the asymptotic properties of GEE estimates using the maxmin method. Xie \& Yang (2003) analyzed the asymptotic properties of GEE in the case of a single covariate, as the number of individuals, the number of observations per individual, or both grow to infinity. Balan \& Schiopu-Kratina (2005) employed pseudo-likelihood equations to demonstrate the existence, weak consistency, and asymptotic normality of GEE estimators when the covariate dimension is ﬁxed. For the analysis of high-dimensional longitudinal data, Wang (2011) extended the asymptotic properties of GEE estimators with binary response variable  when the number of covariates grows to inﬁnity. Wang et al. (2012) consider the penalized GEE for analyzing longitudinal data with high-dimensional covariates.

With the rapid advancement of economic development and information technology, data collection capabilities have significantly improved, leading to a dramatic increase in the scale of longitudinal data. As a consequence, we are facing not only the difficulties associated with high dimensionality but also the computational and storage challenges brought about by the explosive growth in data volume. Take the China Household Finance Survey (CHFS) as an example: since its launch in 2011, it has covered over 40,000 households and conducted follow-up surveys every two years, resulting in millions of observations across thousands of economic and social variables, such as assets, liabilities, and consumption, that capture household financial behavior over time. Similarly, the U.S. National Health and Nutrition Examination Survey (NHANES) has accumulated tens of thousands of longitudinal records with hundreds of health indicators, further illustrating the pressure of rapidly increasing sample sizes alongside high dimensionality. These characteristics render conventional storage and analytical approaches computationally infeasible, highlighting the urgent need for more efficient algorithms, scalable computing frameworks, and distributed processing techniques to effectively handle modern longitudinal data.

To efficiently process such large-scale datasets, several methods have been proposed, including divide-and-conquer strategies (Lin \& Xi 2011; Xu et al. 2020), online updating techniques for streaming data (Luo et al. 2023; Schifano et al. 2016 ), and subsampling approaches (Fithian \& Hastie 2014; Ma et al. 2015; Wang et al. 2018). Among them, subsampling methods have received significant attention for their effectiveness in reducing resource consumption and preserving data representativeness, and have resulted in substantial theoretical and practical achievements. For cross-sectional data, Fithian \& Hastie (2014) introduced a Poisson sampling method in the context of logistic regression. Ma et al. (2015) conducted a statistical analysis of leverage-based subsampling. Wang et al. (2018) derived optimal subsampling probabilities for logistic regression based on the A-optimality criterion, and subsequently proposed a two-step adaptive algorithm aimed at approximating this optimal subsampling scheme. Yu et al. (2022) constructed optimal Poisson subsampling probabilities for pseudo-likelihood estimation, guided by A- and L-optimality criteria, and developed a distributed framework to handle data partitioned across multiple blocks or locations. Yao \& Wang (2019)  and Yao et al. (2023) applied optimal subsampling and Poisson-based subsampling methods for softmax regression models. Ai et al. (2021) applied optimal subsampling methods for generalized linear models. Yuan et al. (2024) incorporated subsampling strategies into distributed composite quantile regression frameworks. For longitudinal data, Wang et al. (2023) developed a new subsampling strategy that incorporates leverage and gradient information. Han \& Fu (2023) developed optimal subsampling algorithms for marginal model.

Recently, for high-dimensional data, Gao et al. (2024) and Shan \& Wang (2024) investigated subsampling strategies based on decorrelated score approaches for generalized linear models. Li et al. (2024) investigated a Poisson-based subsampling method for expectile regression in large-scale data. To the best of our knowledge, no prior work has explored optimal subsampling algorithms in the context of growing-dimensional longitudinal settings. We aim to develop an optimal Poisson subsampling algorithm for GEE with high-dimensional covariates. This study makes the following main contributions: (i) We propose a Poisson subsampling algorithm for GEE in  growing-dimensional longtitual data and establish the consistency and asymptotic normality of the resulting estimator. (ii) We further develop a two-step algorithm aimed at approximating the optimal Poisson sampling probabilities, thereby extracting more informative subsamples for estimation.

The remainder of this paper is organized as follows. In Section 2, we establish the asymptotic properties of the general Poisson subsampling estimator. Section 3 presents the optimal Poisson subsampling probabilities, which are determined according to the A- and L-optimality criteria. Simulation studies in Section 4 demonstrate the effectiveness of the proposed method. In Section 5, we apply the proposed method to CHFS dataset. The proof can be found in the Appendix. 

\section{Poisson subsampling method based on generalized estimating equations} \setcounter {equation}{0}
\def\theequation{\thesection.\arabic{equation}}
\subsection{Generalized estimating equations}

 For $i = 1, \cdots, n$ and $j = 1, \cdots, m_i$, let $y_{ij}$ denote the response variable and $\mathbf{x}_{ij}$ represent a covariate vector with diverging dimension $p_n$. Define $\mathbf{Y}_i = \left( y_{i1}, \cdots, y_{im_i} \right)^T$ and $\mathbf{X}_i = \left( \mathbf{x}_{i1}, \cdots, \mathbf{x}_{im_i} \right)^T$. Without loss of generality, we assume that $m_1 = \cdots = m_n=m$. The conditional expectation of $y_{ij}$ is given by $E\left(y_{ij} \mid \mathbf{x}_{ij}\right) = \mu_{ij}$, where $\mu_{ij}=g\left( \eta_{ij} \right).$ Here, $g\left( \cdot \right)$ is a known link function, $\eta_{ij}=\mathbf{x}_{ij}^T\boldsymbol{\beta}$, and $\boldsymbol{\beta}\in \mathcal{B} \subset R^{p_n}$ is the regression parameter vector. Observations within the same individual are assumed to be correlated, whereas those from different individuals are independent. Let $\boldsymbol{\mu}_{i} = \left( \mu_{i1}, \cdots, \mu_{im} \right)^{T}$ be the marginal mean vector for the $i$-th individual, and the covariance of the response variable $\mathbf{Y}_i$ is given by:
\begin{equation}
\text{cov}\left(\mathbf{Y}_i\right) = \mathbf{V}_i\left(\boldsymbol{\beta}\right)= \phi \mathbf{A}_i^{1/2}\left(\boldsymbol{\beta}\right)\mathbf{R}^{-1}\mathbf{A}_i^{1/2}\left(\boldsymbol{\beta}\right) ,
\end{equation}
where $\mathbf{A}_i = \mathrm{diag}\left(\mathrm{Var}\left(y_{i1}\right), \cdots, \mathrm{Var}\left(y_{im}\right)\right)$ is a diagonal matrix, $\mathbf{R}$ is the true correlation matrix of the response variable $\mathbf{Y}_i$, and $\phi$ is the dispersion parameter, which may be known or unknown.

Liang \& Zeger (1986) introduced the generalized estimating equation, which takes the following form:
\[
\sum_{i=1}^n\mathbf{X}_i^T\mathbf{A}_i\left(\boldsymbol{\beta}\right)\mathbf{V}_i^{-1}\left(\boldsymbol{\beta}\right)\left(\mathbf{Y}_i-\boldsymbol{\mu}_i\left(\boldsymbol{\beta}\right)\right)=0.
\]

We use $\hat{\mathbf{R}}$ to denote the  estimated working correlation matrix, and define the GEE estimator $\hat{\boldsymbol{\beta}}$ by:
\[
\mathbf{S}_n\left(\boldsymbol{\beta}\right)=\frac{1}{n}\sum_{i=1}^n\mathbf{X}_i^T\mathbf{A}_i^{1/2}\left(\boldsymbol{\beta}\right)\hat{\mathbf{R}}^{-1}\mathbf{A}_i^{-1/2}\left(\boldsymbol{\beta}\right)\left(\mathbf{Y}_i-\boldsymbol{\mu}_i\left(\boldsymbol{\beta}\right)\right)=0,
\] 
where $\boldsymbol{\varepsilon}_{i}\left(\boldsymbol{\beta}\right) =\mathbf{A}_i^{-1/2}\left(\boldsymbol{\beta}\right)(\mathbf{Y}_{i} - \boldsymbol{\mu}_{i}\left(\boldsymbol{\beta}\right))$, $\hat{\mathbf{R}}^{(k)}$ and $\hat{\boldsymbol{\beta}}^{(k)}$ represent the values obtained during the $k$-th update. Given $\hat{\mathbf{R}}^{(k)}$ and $\hat{\boldsymbol{\beta}}^{(k-1)}$, the estimator $\hat{\boldsymbol{\beta}}^{(k)}$ can be computed by:
\begin{equation*}
\begin{aligned}
\hat{\boldsymbol{\beta}}^{(k)} = \Bigg( \sum_{i=1}^{n} &\mathbf{X}_i^{T} \mathbf{A}_i^{1/2}(\hat{\boldsymbol{\beta}}^{(k-1)})
\left( \hat{\mathbf{R}}^{(k)} \right)^{-1}
\mathbf{A}_i^{1/2}(\hat{\boldsymbol{\beta}}^{(k-1)}) \mathbf{X}_i \Bigg)^{-1} \\
&\times \Bigg( \sum_{i=1}^{n} \mathbf{X}_i^{T} \mathbf{A}_i^{1/2}(\hat{\boldsymbol{\beta}}^{(k-1)})
\left( \hat{\mathbf{R}}^{(k)} \right)^{-1}
\boldsymbol{\varepsilon}_i(\hat{\boldsymbol{\beta}}^{(k-1)}) \Bigg).
\end{aligned}
\end{equation*}

We repeat the iterative procedure until the norm of the difference between successive estimates of $\boldsymbol{\beta}$ is less than $10^{-4}$. The resulting estimate corresponds to the previously defined $\hat{\boldsymbol{\beta}}$.

To obtain consistent parameter estimates, the computational complexity is at least $O(c \cdot nmp_n^{2})$, where $c$ denotes the number of iterations. As the number of individuals $n$ increases, the computational burden also increases. Typically, subsampling algorithms help to reduce computational costs. The Poisson subsampling method can avoid memory overflow issues while maintaining efficient parameter estimation. Therefore, the next subsection will introduce the Poisson subsampling method for parameter estimation.

\subsection{Poisson subsampling algorithm}
Let $D=\left\{ \left( \mathbf{X}_i^*,\mathbf{Y}_i^*,\pi_i^* \right) \right\} _{i=1}^{r^*}$ represent the subsample dataset, where $\pi_i^*$ denotes the sampling probability for individual $i$. Let $\tilde{\mathbf{R}}$ and $\tilde{\boldsymbol{\beta}}$ represent the estimates based on the subsample; $\tilde{\mathbf{R}}^{(k)}$ and $\tilde{\boldsymbol{\beta}}^{(k)}$ represent the values obtained during the $k$-th update. Given the subsample $D$, the weighted generalized estimating equation takes the following form:
\begin{equation}\label{2} \mathbf{S}_r\left(\boldsymbol{\beta}\right) = \frac{1}{n}\sum_{i=1}^{r^*}\frac{1}{\pi_i^*}\mathbf{X}_i^{*T}\mathbf{A}_i^{*1/2}\left(\boldsymbol{\beta}\right)\tilde{\mathbf{R}}^{-1}\boldsymbol{\varepsilon}_{i}^*(\boldsymbol{\beta})=0. \end{equation}

Under the assumption of a working independence correlation matrix, the initial estimate of $\boldsymbol{\beta}$ can be directly obtained. $\tilde{\mathbf{R}}^{(k)}$ is estimated using the Gaussian pseudo-likelihood method. With $\tilde{\mathbf{R}}^{(k)}$ and $\tilde{\boldsymbol{\beta}}^{(k-1)}$, the value of $\tilde{\boldsymbol{\beta}}^{(k)}$ can be estimated as:
\begin{equation} \label{7}
\begin{aligned}
\tilde{\boldsymbol{\beta}}^{(k)} = \Bigg( \sum_{i=1}^{r^*} &\mathbf{X}_i^{*T}\mathbf{A}_i^{*1/2}(\tilde{\boldsymbol{\beta}}^{(k-1)})
\left(\tilde{\mathbf{R}}^{(k)}\right)^{-1}
\mathbf{A}_i^{*1/2}(\tilde{\boldsymbol{\beta}}^{(k-1)})\mathbf{X}_i^* \Bigg)^{-1} \\
&\times \Bigg( \sum_{i=1}^{r^*} \mathbf{X}_i^{*T} \mathbf{A}_i^{*1/2}(\tilde{\boldsymbol{\beta}}^{(k-1)})
\left(\tilde{\mathbf{R}}^{(k)}\right)^{-1}
\boldsymbol{\varepsilon}_i^*(\tilde{\boldsymbol{\beta}}^{(k-1)}) \Bigg).
\end{aligned}
\end{equation}
Equation (\ref{7}) is iteratively applied until  $\|\tilde{\boldsymbol{\beta}}^{(k+1)}-\tilde{\boldsymbol{\beta}}^{(k)} \|<10^{-4}$. The resulting estimate corresponds to the previously defined $\tilde{\boldsymbol{\beta}}$. We illustrate the steps of a general Poisson subsampling algorithm in Algorithm 1. 

\paragraph*{Algorithm 1} General Poisson Subsampling Algorithm.
\begin{itemize}
	\item \textbf{Step 1:} Initialize the set $\mathcal{D} = \emptyset$. For each $i = 1, \dots, n$, generate an independent Bernoulli random variable $\delta_i \sim \text{Bernoulli}(\pi_i)$. If $\delta_i = 1$, include the triplet $(\mathbf{X}_i, \mathbf{Y}_i, \pi_i)$ in the set $\mathcal{D}$.
 
	\item \textbf{Step 2:} Based on the subsample $\mathcal{D}$, we use equation \eqref{7} to estimate the weighted generalized estimating equation in \eqref{2} and obtain the regression parameter estimate $\tilde{\boldsymbol{\beta}}$.
\end{itemize}

Whether the observation $(\mathbf{X}_i^*, \mathbf{Y}_i^*)$ of individual $i$ is included in the subsample depends solely on its own probability $\pi_i$, without considering the sampling probabilities of other individuals. In Algorithm 1, a random variable is generated through a Bernoulli trial to decide whether the observation $(\mathbf{X}_i^*, \mathbf{Y}_i^*)$ of individual $i$ is included in the subsample. Therefore, for massive data sets, the Poisson subsampling method alleviates memory constraint issues.

The size of the drawn subsample, denoted by $r^*$, satisfies $E(r^*) = \sum_{i=1}^n \pi_i$. Let $r = \sum_{i=1}^n \pi_i$ denote the expected size of the drawn subsample. Furthermore, assume that $r \ll n$, which is a common assumption in big data scenarios. The following regularity conditions are imposed to ensure consistency and asymptotic normality.
\begin{itemize}
\item[(C1)] $\sup_{i,j}\left\|\mathbf{X}_{ij}\right\|=O\left(\sqrt{p_n}\right)$.
\item[(C2)] The parameter vector $\boldsymbol{\beta}$ is assumed to lie within a compact set $\mathcal{B} \subseteq \mathbb{R}^{p_n}$, and the true parameter vector $\boldsymbol{\beta}_0$ is also contained in $\mathcal{B}$.

\item[(C3)] There are positive constants $b_1, b_2>0$ such that $$b_1\leq\lambda_{\min}\left(\frac{1}{n}\sum_{i=1}^n\mathbf{X}_i^T\mathbf{X}_i\right)\leq\lambda_{\max}\left(\frac{1}{n}\sum_{i=1}^n\mathbf{X}_i^T\mathbf{X}_i\right)\leq b_2,$$ and similarly, two other constants $b_3, b_4>0$ satisfy $$b_3\leq\lambda_{\min}\left(\frac{r}{n}\sum_{i=1}^n\frac{1}{n\pi_i}\mathbf{X}_i^T\mathbf{X}\right)\leq\lambda_{\max}\left(\frac{r}{n}\sum_{i=1}^n\frac{1}{n\pi_i}\mathbf{X}_i^T\mathbf{X}_i\right)\leq b_4,$$ where $\lambda_{\min}$ and $\lambda_{\max}$ indicate the smallest and largest eigenvalues of a matrix, respectively.

\item[(C4)] The true correlation matrix $\mathbf{R}_{0}$ is assumed to have eigenvalues bounded away from 0 and $+\infty$. The estimated working correlation matrix $\tilde{\mathbf{R}}$ satisfies $\|\tilde{\mathbf{R}}^{-1}-\bar{\mathbf{R}}^{-1}\|=O_p(\sqrt{{p_n}/{r}})$, where $\bar{\mathbf{R}}$ is a positive definite matrix with eigenvalues also bounded away from 0 and $+\infty$. We do not require $\bar{\mathbf{R}}$ to be the true working correlation matrix $\mathbf{R}_{0}$.

\item [(C5)] There are positive constants $M_{1}$ and $\delta$ such that $$E\left(\left\|\mathbf{A}_i^{-1/2}\left(\boldsymbol{\beta}\right)\left(\mathbf{Y}_i-\boldsymbol{\mu}_i\left(\boldsymbol{\beta}\right)\right)\right\|^{2+\delta}\right)\leq M_1.$$

\item [(C6)] Let $M_{2}>0$ be a constant satisfying $0\leq\dot{\boldsymbol{\mu}}(\mathbf{X}_{ij}^T\boldsymbol{\beta})\leq\infty$, and $0\leq\ddot{\boldsymbol{\mu}}(\mathbf{X}_{ij}^T\boldsymbol{\beta}),\dddot{\boldsymbol{\mu}}(\mathbf{X}_{ij}^T\boldsymbol{\beta})\leq M_2$, where $\dot{\boldsymbol{\mu}}(\mathbf{X}_{ij}^T\boldsymbol{\beta}),\ddot{\boldsymbol{\mu}}(\mathbf{X}_{ij}^T\boldsymbol{\beta}),\dddot{\boldsymbol{\mu}}(\mathbf{X}_{ij}^T\boldsymbol{\beta})$ are the first-, second-, and third-order derivatives of $\boldsymbol{\mu}(\mathbf{X}_{ij}^T\boldsymbol{\beta})$, respectively.

\item [(C7)] $\max_{1\leq i\leq n}\frac{1}{n\pi_i}=O_p\left(\frac{1}{r}\right)$.
\end{itemize}

Assumption (C1) is a common condition for diverging-dimension M-estimators, which aligns with the setting in Portnoy(1985). Assumption (C2) is a necessary condition for the consistency of the estimator and has been widely adopted in many studies, including Newey \& McFadden(1994). Assumption (C3) is frequently used in high-dimensional regression literature, with a similar formulation appearing in Wang (2011). Assumption (C4) extends the framework from fixed $p_n$ to high-dimensional $p_n$. Assumption (C5) imposes moment conditions on the model. Assumption (C6) ensures the consistency of parameter estimation. Assumption (C7) restricts the weights in the weighted generalized estimating equations, primarily to prevent individuals with extremely small subsampling probabilities from unduly influencing the results.

\begin{theorem}\label{theorem1}
	Under assumptions (C1)-(C7), if ${p_n^2}/{r}=o(1)$ and $\tilde{\boldsymbol{\beta}}$ is a solution to $\mathbf{S}_r(\boldsymbol{\beta})=0$, then
$$||\tilde{\boldsymbol{\beta}}-\boldsymbol{\beta}_{0}||=O_p(\sqrt{p_n/r}).$$
\end{theorem}
\begin{theorem}\label{theorem2}
	Under assumptions (C1)-(C7), if ${p_n^3}/{r}=o(1)$, then for any $\mathbf{c}_n \in \mathbf{R}^{p_n}$ with $\|\mathbf{c}_n\| = 1$, we have
$$\mathbf{c}_n^T\bar{\mathbf{M}}_r^{-1/2}\left(\boldsymbol{\beta}_{0}\right)\bar{\mathbf{H}}_n\left(\boldsymbol{\beta}_{0}\right)(\tilde{\boldsymbol{\beta}}-\boldsymbol{\beta}_{0})\xrightarrow{d}N\left(0,1\right),$$
where 
\begin{align*}
&\bar{\mathbf{M}}_r\left(\boldsymbol{\beta}_{0}\right)=\frac{1}{n^2}\sum_{i=1}^n\frac{1}{\pi_i}\mathbf{X}_i^T\mathbf{A}_i^{1/2}\left(\boldsymbol{\beta}_{0}\right)\bar{\mathbf{R}}^{-1}\boldsymbol{\varepsilon}_i\left(\boldsymbol{\beta}_{0}\right)\boldsymbol{\varepsilon}_i^T\left(\boldsymbol{\beta}_{0}\right)\bar{\mathbf{R}}^{-1}\mathbf{A}_i^{1/2}\left(\boldsymbol{\beta}_{0}\right)\mathbf{X}_i, \\
&\bar{\mathbf{H}}_{n}\left(\boldsymbol{\beta}_{0}\right)=\frac{1}{n}\sum_{i=1}^{n}\mathbf{X}_{i}^{T}\mathbf{A}_{i}^{1/2}\left(\boldsymbol{\beta}_{0}\right)\bar{\mathbf{R}}^{-1}\mathbf{A}_{i}^{1/2}\left(\boldsymbol{\beta}_{0}\right)\mathbf{X}_{i}.
\end{align*}
\end{theorem}

\begin{theorem}\label{theorem3}
	Under assumptions (C1)-(C7), if ${p_n^3}/{r}=o(1)$, then
$$\mathbf{c}_n \tilde{\boldsymbol{\Sigma}} \mathbf{c}_n^T - \mathbf{c}_n \boldsymbol{\Sigma} \mathbf{c}_n^T = o_p(1),$$
where $\mathbf{c}_n$ is a $p_n$-dimensional vector satisfying $\mathbf{c}_n \mathbf{c}_n^T = 1$; 
$\tilde{\boldsymbol{\Sigma}} = \mathbf{H}_n^{-1}(\tilde{\boldsymbol{\beta}})\bar{\mathbf{M}}_r(\tilde{\boldsymbol{\beta}})\mathbf{H}_n^{-1}(\tilde{\boldsymbol{\beta}})$; 
$\boldsymbol{\Sigma} = \bar{\mathbf{H}}_n^{-1}(\boldsymbol{\beta}_{0})\bar{\mathbf{M}}_r(\boldsymbol{\beta}_{0})\bar{\mathbf{H}}_n^{-1}(\boldsymbol{\beta}_{0})$, 
$\mathbf{H}_n(\boldsymbol{\beta})$ is similar to $\bar{\mathbf{H}}_n(\boldsymbol{\beta})$, where $\tilde{\mathbf{R}}$ is used in place of $\bar{\mathbf{R}}$.
\end{theorem}

Theorem \ref{theorem2} indicates that the estimation error of $\tilde{\boldsymbol{\beta}} - \boldsymbol{\beta}_{0}$ follows an asymptotic normal distribution, and its asymptotic distribution is related to the sampling probability $\pi = \begin{Bmatrix} \pi_{i} \end{Bmatrix}_{i=1}^{n}$. Regardless of the correctness of the working correlation matrix specification, increasing the sample size or the number of individual observations will lead to better estimation results for the regression parameter estimator.

\section{Optimal Poisson Sampling Algorithm}
\setcounter{equation}{0}\def\theequation{\thesection.\arabic{equation}}
\subsection{Optimal Poisson Sampling Strategy}
To obtain the regression parameter estimator $\tilde{\boldsymbol{\beta}}$ using \eqref{2}, the Poisson sampling probabilities $\pi = \left\{\pi_i\right\}_{i=1}^n$ need to be specified. The optimal subsampling probabilities can be determined by minimizing $tr(\boldsymbol{\Sigma})$, that is, by using the A-optimality criterion.
\begin{theorem}\label{tthree}
Definition
$h_{i}^{MV} = \left\|\bar{\mathbf{H}}_{n}^{-1}\left(\boldsymbol{\beta}_{0}\right)\mathbf{X}_i^T\mathbf{A}_i^{1/2}\left(\boldsymbol{\beta}_{0}\right)\bar{\mathbf{R}}^{-1}\boldsymbol{\varepsilon}_i\left(\boldsymbol{\beta}_{0}\right)\right\|$, $i=1,2,\cdots,n$, let $h_{(1)}^{MV} \leq h_{(2)}^{MV} \leq \cdots \leq h_{(n)}^{MV}$ be the order statistics of $\left\{h_i^{MV}\right\}_{i=1}^n$. If the subsampling probability is
\begin{equation}
\begin{aligned}\label{11}
\pi_{i}^{MV} & = r \frac{h_{i}^{MV} \wedge T}{\sum_{j=1}^{n} h_{j}^{MV} \wedge T},
\end{aligned}
\end{equation}
the value of $tr(\boldsymbol{\Sigma})$ is minimized, where,
\[
T = \sum_{i=1}^{n-w} h_{(i)}^{MV} / (r-w),
\]
and
\[
w = \min\left\{s \mid 0 \leq s \leq r, h_{(n-s)}^{MV} < \sum_{i=1}^{n-s} h_{(i)}^{MV} / (r-s)\right\}.
\]
\end{theorem}

The computation of $T$ in \eqref{11} is required only for those individuals $i$ satisfying the condition $r h_{(i)}^{MV} / \sum_{j=1}^n h_j^{MV} > 1$. In this case, $s$ is the number of individuals for which $p_{i}^{MV} = 1$. If all individuals satisfy $rh_{(i)}^{MV}/\sum_{j=1}^n (h_j^{MV}) \leq 1$, then we can directly set $p_i^{MV} = rh_{(i)}^{MV}/\sum_{j=1}^n (h_j^{MV})$. To reduce computational complexity, the next theorem establishes the L-optimal subsampling strategy.
\begin{theorem}\label{tfour}
	Definition $h_{i}^{MVc} = \left\|\mathbf{X}_i^T\mathbf{A}_i^{1/2}\left(\boldsymbol{\beta}_{0}\right)\bar{\mathbf{R}}^{-1}\boldsymbol{\varepsilon}_i\left(\boldsymbol{\beta}_{0}\right)\right\|$, $i=1,2,\cdots,n$, let $h_{(1)}^{MVc} \leq h_{(2)}^{MVc} \leq \cdots \leq h_{(n)}^{MVc}$ be the order statistics of $\left\{h_i^{MVc}\right\}_{i=1}^n$. If the subsampling probability is
\begin{equation}
\begin{aligned}
\pi_{i}^{MVc} & = r \frac{h_{i}^{MVc} \wedge T}{\sum_{j=1}^{n} h_{j}^{MVc} \wedge T},
\end{aligned}
\end{equation}
then $tr(\bar{\mathbf{M}}_r\left(\boldsymbol{\beta}_{0}\right))$ is minimized, where,
\[
T = \sum_{i=1}^{n-w} h_{(i)}^{MVc} / (r-w),
\]
and
\[
w = \min\left\{s \mid 0 \leq s \leq r, h_{(n-s)}^{MVc} < \sum_{i=1}^{n-s} h_{(i)}^{MVc} / (r-s)\right\}.
\]
\end{theorem}

\subsection{Two-Step Algorithm}
To simplify notation, let $\pi_i^{os}$ and $h_i^{os}$ denote $\pi_i^{MV}$ or $\pi_i^{MVc}$, and $h_i^{MV}$ or $h_i^{MVc}$, respectively, as defined in Theorems \ref{tthree} and \ref{tfour}. Since the computation of $h_{i}^{os}$ is related to the true parameter $\boldsymbol{\beta}_{0}$, the optimal Poisson subsampling probability is
\begin{equation}
\begin{aligned}
\pi_{i}^{os} & = r \frac{h_{i}^{os} \wedge T}{\sum_{i=1}^{n} h_{i}^{os} \wedge T} = r \frac{h_{i}^{os} \wedge T}{n\Psi}, \quad i=1,\cdots,n,
\end{aligned}
\end{equation}
which cannot be directly computed, where $\Psi = n^{-1} \sum_{i=1}^{n} h_i^{os} \wedge T$. To implement this procedure, a two-step algorithm is proposed. In the first step, a pilot subsample of expected size \( r_1 \) is drawn through uniform Poisson sampling, denoted as 
$D_{r_1^*} = \left\{ \left( \mathbf{X}_i^*, \mathbf{Y}_i^*, r_1/n \right) \right\}_{i=1}^{r_1^*}$. Based on \( D_{r_{1}^{*}} \), and assuming an independent working correlation matrix, the resulting estimate \(\tilde{\boldsymbol{\beta}}_{r_1^*}\) is used as an initial approximation of \( \boldsymbol{\beta}_{0}\). The working correlation matrix $\bar{\mathbf{R}}$ is accordingly replaced by $\tilde{\mathbf{R}}_{r_1^*}$, computed via the Gaussian pseudo-likelihood method. Next, the values of $T$ and $\Psi$ can be computed. The purpose of $T$ is to control $\max_{1 \leq i \leq n} \pi_i^{os} = 1$, and since $r \ll n$ is a common case, this implies that the situation where $h_i^{os} > T$ is very rare. Therefore, with sufficiently small subsampling rates, directly setting $T = \infty$ performs quite well. For the estimation of $\Psi$, it can be calculated as $\hat{\Psi} = (r_{1}^{*})^{-1} \sum_{i \in D_{r_1^*}} \left\| \mathbf{L}\mathbf{H}_{r_1}^{-1}(\tilde{\boldsymbol{\beta}}_{r_1^*})\mathbf{X}_i^{*T}\mathbf{A}_i^{*1/2}(\tilde{\boldsymbol{\beta}}_{r_1^*})\tilde{\mathbf{R}}_{r_1^*}^{-1}\boldsymbol{\varepsilon}_i^*(\tilde{\boldsymbol{\beta}}_{r_{1}^{*}})\right\|$, where \(\mathbf{H}_{r_1}(\tilde{\boldsymbol{\beta}}_{r_1^*})=(r_{1}^{*})^{-1} \sum_{i \in D_{r_1^*}}\mathbf{X}_{i}^{*T}\mathbf{A}_{i}^{*1/2}(\tilde{\boldsymbol{\beta}}_{r_1^*})\tilde{\mathbf{R}}^{-1}\mathbf{A}_{i}^{*1/2}(\tilde{\boldsymbol{\beta}}_{r_1^*})\mathbf{X}_{i}^*\). When $\mathbf{L} = \mathbf{I}$, this corresponds to the A-optimality criterion, when $\mathbf{L} = \mathbf{H}_{r_1}(\tilde{\boldsymbol{\beta}}_{r_1^*})$, this corresponds to the L-optimality criterion. Therefore, the optimal subsampling probability can be approximated by $\hat{\pi}_i^{os}$, where $\tilde{\boldsymbol{\beta}}_{r_1^*}$, $\tilde{\mathbf{R}}_{r_1^*}$, $T = \infty$, $\mathbf{H}_{r_1}(\boldsymbol{\beta}_{r_1^*})$, and $\hat{\Psi}$ are used in place of $\boldsymbol{\beta}_{0}$, $\bar{\mathbf{R}}$, $T$, $\bar{\mathbf{H}}_{n}\left(\boldsymbol{\beta}_{0}\right)$, and $\Psi$.

To enhance the robustness of the estimator, we employ the shrinkage-based subsampling method studied by Ma et al. (2015), which combines the optimal subsampling probability $\hat{\pi}_i^{os}$ with the uniform probability:
\begin{equation}\label{9}
\hat{\pi}_i^{sos} = ( 1 - \rho ) \frac{r_2 \left\| \mathbf{L}\mathbf{H}_{r_1}^{-1}(\tilde{\boldsymbol{\beta}}_{r_1^*})\mathbf{X}_i^T\mathbf{A}_i^{1/2}(\tilde{\boldsymbol{\beta}}_{r_1^*})\tilde{\mathbf{R}}_{r_1^*}^{-1}
\boldsymbol{\varepsilon}_i(\tilde{\boldsymbol{\beta}}_{r_1^*})\right\|}{n \hat{\Psi}} + \rho \frac{r_2}{n},
\end{equation}
where $\rho \in (0,1)$, and $r_2$ denotes the expected subsample size in the second step. In practice, it is possible that $\hat{\pi}_i^{sos}$ exceeds 1 due to the shrinkage adjustment. Therefore, the final subsampling probability is given by $\left( \hat{\pi}_i^{sos} \wedge 1 \right)$, and the final regression parameter estimator is denoted as $\breve{\boldsymbol{\beta}}$. Algorithm 2 provides a full description of the two-step algorithm. The asymptotic properties of the regression parameter $\breve{\boldsymbol{\beta}}$ obtained from Algorithm 2 are presented in Theorem \ref{tfive}.
\paragraph*{Algorithm 2} Two-Step Algorithm.
\begin{itemize}
	\item \textbf{Step 1:} Use the uniform Poisson subsampling probability $\{\pi_i = r_1/n\}_{i=1}^n$ to draw the pilot sample $D_{r_{1}^{*}}$ and obtain $\tilde{\boldsymbol{\beta}}_{r_1^*}$, $\tilde{\mathbf{R}}_{r_1^*}$, $T = \infty$, $\mathbf{H}_{r_1}(\boldsymbol{\beta}_{r_1^*})$, and $\hat{\Psi}$. Compute the optimal Poisson subsampling probability $\left(\hat{\pi}_{i}^{sos} \wedge 1\right)$ based on \eqref{9}.
	\item \textbf{Step 2:} With the approximate optimal subsampling probabilities obtained from Step 1, draw the sample $D_{r_{2}^{*}}$, and perform regression parameter estimation.
\end{itemize}
\begin{theorem}\label{tfive}
    Under assumptions (C1)-(C7), if ${p_n^3}/{r}=o(1)$ and the condition $r_1 r_2^{-1/2}=o(1)$, then for any $\mathbf{c}_n \in \mathbf{R}^{p_n}$ with $\|\mathbf{c}_n\| = 1$, we have
\[
\mathbf{c}_n^T\left( \bar{\mathbf{H}}_n^{-1}(\boldsymbol{\beta}_{0})\bar{\mathbf{M}}^L_r(\boldsymbol{\beta}_{0})\bar{\mathbf{H}}_n^{-1}(\boldsymbol{\beta}_{0}) \right)^{-1/2} ( \breve{\boldsymbol{\beta}} -\boldsymbol{\beta}_{0})\xrightarrow{d} N(0, 1),
\]
where
\[
\bar{\mathbf{M}}^L_r(\boldsymbol{\beta}_{0}) = \frac{1}{n^2} \sum_{i=1}^n \frac{\mathbf{X}_i^T\mathbf{A}_i^{1/2}\left(\boldsymbol{\beta}_{0}\right)\bar{\mathbf{R}}^{-1}\boldsymbol{\varepsilon}_i\left(\boldsymbol{\beta}_{0}\right)\boldsymbol{\varepsilon}_i^T\left(\boldsymbol{\beta}_{0}\right)\bar{\mathbf{R}}^{-1}\mathbf{A}_i^{1/2}\left(\boldsymbol{\beta}_{0}\right)\mathbf{X}_i}{\pi_i^{sos} \wedge 1},
\]
and
\[
\pi_i^{sos} = \left( 1 - \rho \right) \frac{r_2 \left\| \mathbf{L} \bar{\mathbf{H}}_{n}^{-1}\left(\boldsymbol{\beta}_{0}\right)\mathbf{X}_i^T\mathbf{A}_i^{1/2}\left(\boldsymbol{\beta}_{0}\right)\bar{\mathbf{R}}^{-1}\boldsymbol{\varepsilon}_i\left(\boldsymbol{\beta}_{0}\right)
\right\|}{\sum_{i=1}^n \left\| \mathbf{L} \bar{\mathbf{H}}_{n}^{-1}\left(\boldsymbol{\beta}_{0}\right)\mathbf{X}_i^T\mathbf{A}_i^{1/2}\left(\boldsymbol{\beta}_{0}\right)\bar{\mathbf{R}}^{-1}\boldsymbol{\varepsilon}_i\left(\boldsymbol{\beta}_{0}\right)\right\|} + \rho \frac{r_2}{n}.
\]
\end{theorem}

According to Theorem \ref{tfive}, the covariance matrix of $\breve{\boldsymbol{\beta}}$ can be estimated as $\mathbf{H}_{r_1}^{-1}(\breve{\boldsymbol{\beta}})\breve{\mathbf{M}}_r(\breve{\boldsymbol{\beta}})
\\
\mathbf{H}_{r_1}^{-1}(\breve{\boldsymbol{\beta}})$, where $\mathbf{H}_{r_1}(\breve{\boldsymbol{\beta}}) = n^{-1} \sum_{i \in D_{r_{2}^{*}}}\mathbf{X}_{i}^{*T}\mathbf{A}_{i}^{*1/2}(\breve{\boldsymbol{\beta}})\tilde{\mathbf{R}}_{r_1^*}^{-1}\mathbf{A}_{i}^{*1/2}(\breve{\boldsymbol{\beta}})\mathbf{X}_{i}^*,$
and
$\breve{\mathbf{M}}_r(\boldsymbol{\beta}_{r_1^*}) = n^{-2} \sum_{i \in D_{r_{2}^{*}}} 
\\
\left( \hat{\pi}_i^{*sos} \wedge 1 \right)^{-1}\mathbf{X}_i^{*T}\mathbf{A}_i^{*1/2}(\breve{\boldsymbol{\beta}})\tilde{\mathbf{R}}_{r_1^*}^{-1}\boldsymbol{\varepsilon}_i^*(\breve{\boldsymbol{\beta}})\boldsymbol{\varepsilon}_i^{*T}(\breve{\boldsymbol{\beta}})\tilde{\mathbf{R}}_{r_1^*}^{-1}\mathbf{A}_i^{*1/2}(\breve{\boldsymbol{\beta}})\mathbf{X}_i^*.$

\section{Numerical Simulation}
\setcounter{equation}{0}\def\theequation{\thesection.\arabic{equation}} 
We assess the effectiveness of the optimal Poisson subsampling algorithm through simulation studies, considering a linear regression model in the context of high-dimensional longitudinal data:
\[\mathbf{Y}_i = \mathbf{X}_i \boldsymbol{\beta} + \boldsymbol{\varepsilon}_i, \quad i = 1, 2, \cdots, n.\]
The true value  $\boldsymbol{\beta}_0 = (1, 1.5, 1, 1.5, \cdots, 1, 1.5)^T$. We consider three settings for the dimensionality: $p_n$ = 30, 50, and 70. The covariates $\mathbf{X}_i$ are generated from two different distributions:

\begin{itemize}
\item[(1)] \textbf{Case 1:} $\mathbf{X}_i$ follows a multivariate $t$-distribution with $3$ degrees of freedom, i.e., $t_3(\mathbf{0}, \boldsymbol{\Sigma})$, where $\boldsymbol{\Sigma} = (0.5^{|i-j|})$.
\item[(2)] \textbf{Case 2:} $\mathbf{X}_i$ follows a log-normal distribution $LN(\mathbf{0},\, 1.8\boldsymbol{\Sigma})$.
\end{itemize}

The error term $\boldsymbol{\varepsilon}_i$ is generated from a multivariate normal distribution $N\left(\mathbf{0}, \mathbf{R}(\boldsymbol{\alpha})\right)$, where the correlation parameter is set to $\alpha = 0.5$. We consider three different work correlation matrices : EX, AR(1), MA(1).

The number of observation individuals is set to 10000, with each individual having 5 observations, $m=5$. The pilot subsample size is set to $r_1 = 200$, and the second-stage subsample size $r_2$ is chosen from {100, 200, 400, 600, 800, 1000}. The mean squared error (MSE) is calculated across 1000 simulation replications to assess the performance of the subsampling methods. Here, $\tilde{\boldsymbol{\beta}}^{(s)}$ denotes the estimated parameter obtained from the $s$-th subsample,
\[\operatorname{MSE} = \frac{1}{1000} \sum_{s=1}^{1000} \left\| \boldsymbol{\beta}_{0} - \tilde{\boldsymbol{\beta}}^{(s)} \right\|^2.\]

\begin{figure}
	\centering
	\includegraphics[width=1\linewidth]{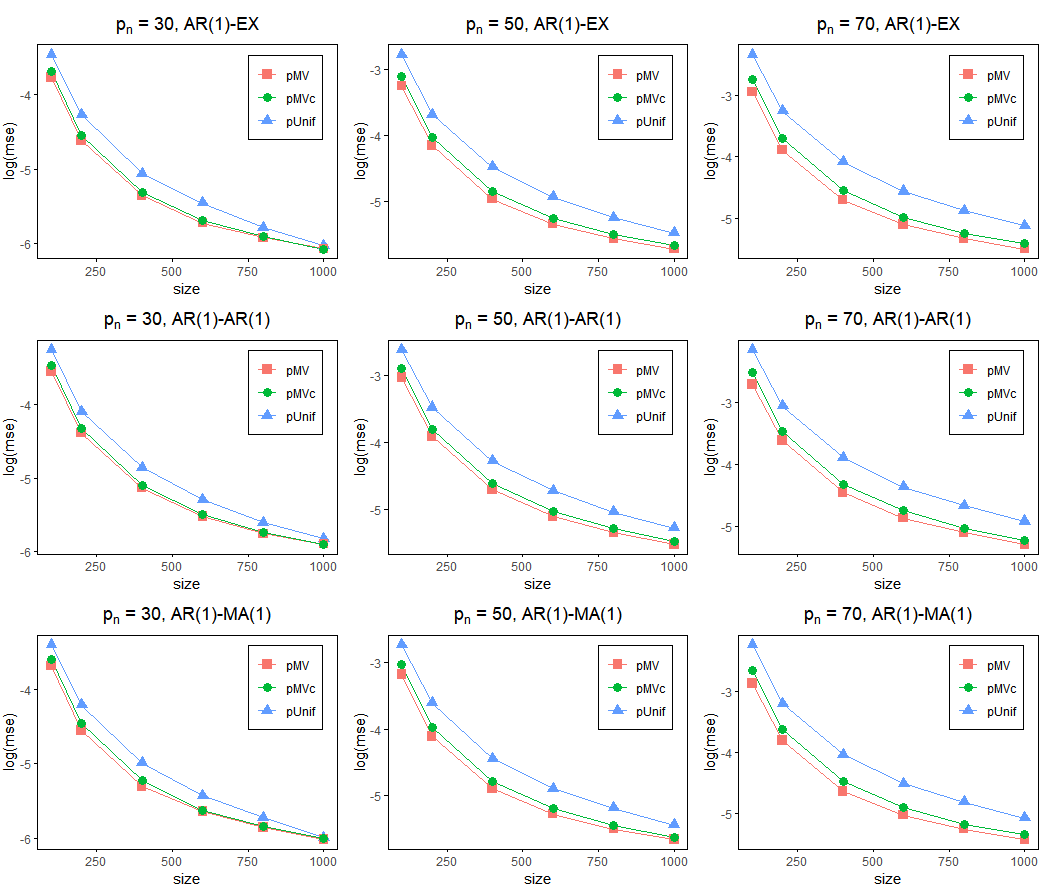}
	\caption{Log(MSE) of the estimator for $p_n$ = 30, 50, and 70 under Case 1 with the true correlation matrix AR(1), where pUnif denotes uniform Poisson subsampling and pMV and pMVc denote optimal Poisson subsampling based on the A- and L-criteria, respectively.}
\end{figure}
\begin{figure}
	\centering
	\includegraphics[width=1\linewidth]{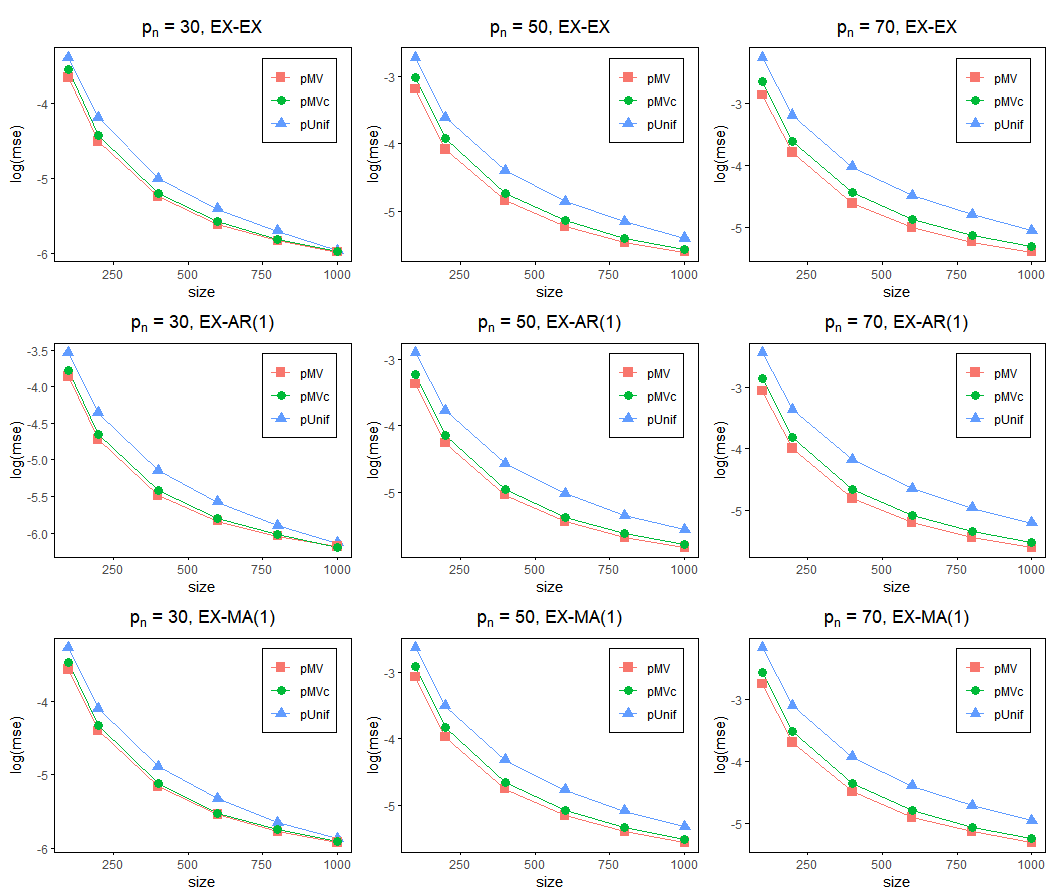}
	\caption{Log(MSE) of the estimator for $p_n$ = 30, 50, and 70 under Case 1 with the true correlation matrix EX, where pUnif denotes uniform Poisson subsampling and pMV and pMVc denote optimal Poisson subsampling based on the A- and L-criteria, respectively.
}
\end{figure}
\begin{figure}
	\centering
	\includegraphics[width=1\linewidth]{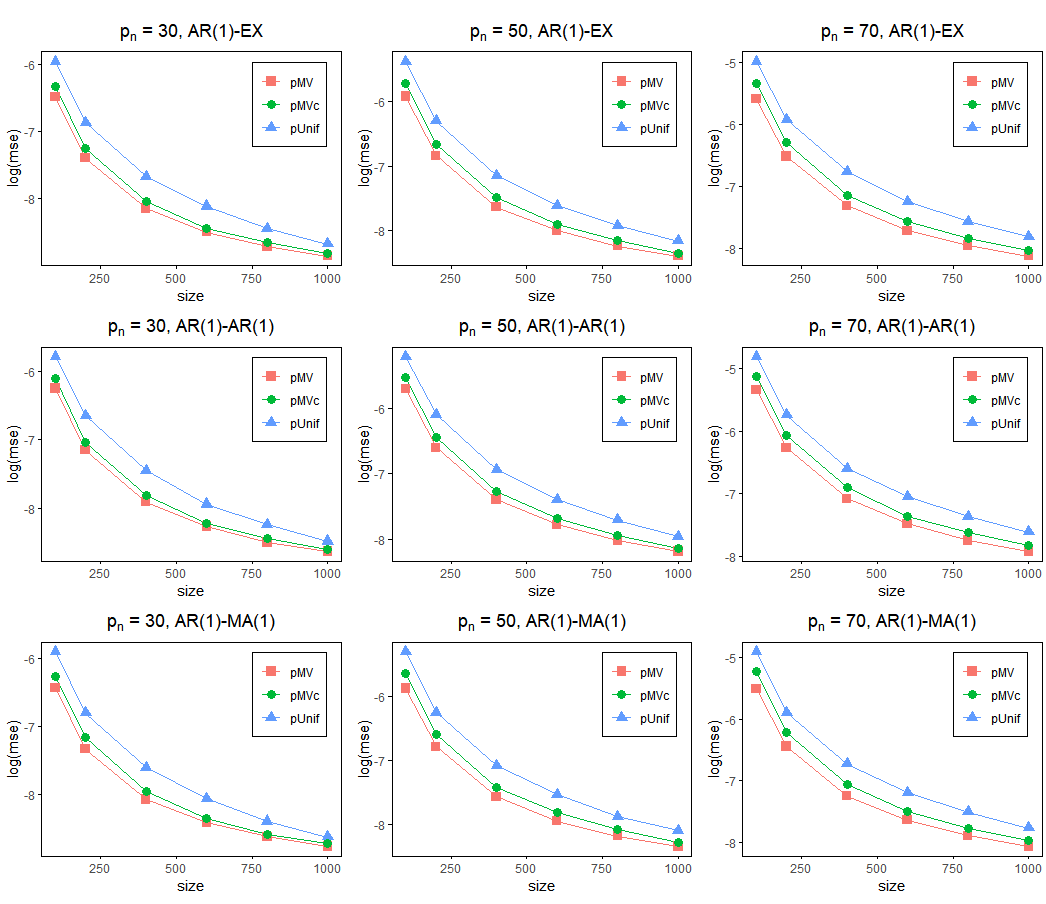}
	\caption{Log(MSE) of the estimator for $p_n$ = 30, 50, and 70 under Case 2 with the true correlation matrix AR(1), where pUnif denotes uniform Poisson subsampling and pMV and pMVc denote optimal Poisson subsampling based on the A- and L-criteria, respectively.}
\end{figure}
\begin{figure}
	\centering
	\includegraphics[width=1\linewidth]{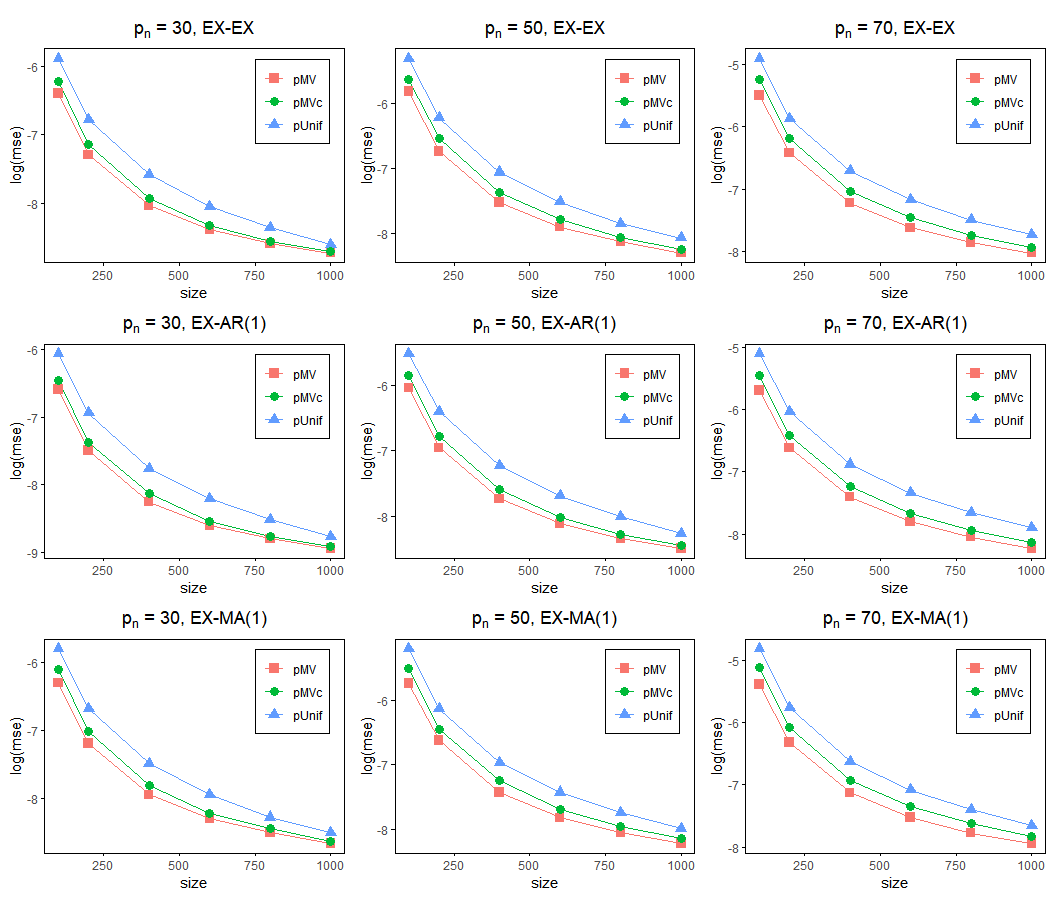}
	\caption{Log(MSE) of the estimator for $p_n$ = 30, 50, and 70 under Case 2 with the true correlation matrix EX, where pUnif denotes uniform Poisson subsampling and pMV and pMVc denote optimal Poisson subsampling based on the A- and L-criteria, respectively.}
\end{figure}

We present the experimental results under Case 1, with AR(1) and EX as the true correlation structures, in Figures 1-2. For example, AR(1)-EX indicates that the left side shows the true correlation matrix, while the right side denotes the working correlation matrix. The uniform Poisson subsampling method is denoted as pUnif, the A-optimal Poisson subsampling as pMV, and the L-optimal Poisson subsampling as pMVc.

The results indicate that as the subsample size increases, the log(MSE) of all three methods decreases gradually. Both pMV and pMVc outperform pUnif significantly, with pMV consistently performing slightly better than pMVc. Even when the working correlation matrix is misspecified, the proposed methods still demonstrate robust performance. We display the experimental results under Case 2, with AR(1) and EX as the true correlation matrices, in Figures 3–4. Similar to the results in Case 1, both pMV and pMVc significantly outperform the uniform Poisson subsampling method.

We summarize the computation time under Case 1 for true correlation matrix AR(1) with $p_n = 30$ and $p_n = 50$ in Table 1. The results show that the computation times for the three Poisson subsampling methods increase with the sample size, but all remain smaller than the computation time using the full dataset. Among them, pMVc exhibits a lower computational cost compared to pMV.

\begin{table*}[!t]
\centering
\caption{Computation time under case 1 with true correlation matrix AR(1) at different dimensions.\label{tab:1}}
\begin{tabular*}{\textwidth}{@{\extracolsep\fill}cccccccc@{\extracolsep\fill}}
\toprule
\multirow{2}{*}{$r_{2}$} & \multirow{2}{*}{Method} & \multicolumn{3}{c}{$p_n=30$} & \multicolumn{3}{c}{$p_n=50$} \\
& & EX & AR(1) & MA(1) & EX & AR(1) & MA(1) \\
\midrule
100  & pUnif & 0.206 & 0.222 & 0.529 & 0.250 & 0.273 & 0.775 \\
     & pMV   & 0.713 & 0.699 & 1.088 & 0.814 & 0.827 & 1.277 \\
     & pMVc  & 0.657 & 0.642 & 0.984 & 0.706 & 0.715 & 1.151 \\
\midrule
200  & pUnif & 0.354 & 0.345 & 0.790 & 0.402 & 0.450 & 1.001 \\
     & pMV   & 0.884 & 0.864 & 1.370 & 0.980 & 0.976 & 1.640 \\
     & pMVc  & 0.804 & 0.781 & 1.281 & 0.844 & 0.862 & 1.469 \\
\midrule
400  & pUnif & 0.686 & 0.683 & 1.432 & 0.713 & 0.710 & 1.604 \\
     & pMV   & 1.268 & 1.234 & 2.058 & 1.420 & 1.384 & 2.504 \\
     & pMVc  & 1.176 & 1.159 & 1.946 & 1.280 & 1.232 & 2.241 \\
\midrule
600  & pUnif & 1.073 & 1.050 & 2.127 & 1.102 & 1.051 & 2.229 \\
     & pMV   & 1.681 & 1.686 & 2.629 & 1.936 & 1.861 & 3.488 \\
     & pMVc  & 1.588 & 1.555 & 2.494 & 1.772 & 1.676 & 3.134 \\
\midrule
800  & pUnif & 1.470 & 1.384 & 2.688 & 1.518 & 1.492 & 3.088 \\
     & pMV   & 2.148 & 2.151 & 3.218 & 2.469 & 2.399 & 4.450 \\
     & pMVc  & 2.005 & 1.991 & 3.038 & 2.249 & 2.195 & 4.007 \\
\midrule
1000 & pUnif & 1.890 & 1.827 & 3.298 & 1.985 & 1.868 & 4.069 \\
     & pMV   & 2.636 & 2.583 & 3.923 & 3.019 & 2.886 & 5.253 \\
     & pMVc  & 2.482 & 2.489 & 3.731 & 2.787 & 2.633 & 4.845 \\
\midrule
full time &     & 15.427 & 15.917 & 22.448 & 15.593 & 16.109 & 22.629 \\
\bottomrule
\end{tabular*}
\end{table*}

\section{Actual Data Analysis}
\setcounter {equation}{0}
\def\theequation{\thesection.\arabic{equation}}
The empirical analysis in this paper is based on data from the China Household Finance Survey (CHFS), which is administered by Southwest University of Finance and Economics (https://chfs.swufe.edu.cn). We perform the analysis at the household level, using total household income (Total\_income) as the response variable. The independent variables include the household head’s residence area (Rural), age (Age), gender (Gender), marital status (Marry), education level (Edu), health status (Unhealth), pension insurance coverage (Endowment\_insurance) and medical insurance coverage (Medinsurance); as well as household-level variables such as total number of family members (Familynum), number of unhealthy family members (Unhealthnum), utility expenditures (Expenditure1), consumption expenditures (Expenditure2), and financial assets (Finanasset). The dataset comprises 9,753 households, with three repeated observations per household.  

The corresponding model is as follows:
\begin{align*}
\text{Total\_income}_{ij} &= \beta_{0} + \beta_{1}\text{Rural}_{ij} + \beta_{2}\text{Age}_{ij} + \beta_{3}\text{Gender}_{ij} + \beta_{4}\text{Marry}_{ij}
+ \beta_{5}\text{Edu}_{ij}  \\
&\quad + \beta_{6}\text{Unhealth}_{ij} + \beta_{7}\text{Familynum}_{ij} + \beta_{8}\text{Unhealthnum}_{ij} \\
&\quad + \beta_{9}\text{Endowment\_insurance}_{ij} + \beta_{10}\text{Medinsurance}_{ij} + \beta_{11}\text{Expenditure1}_{ij} \\
&\quad + \beta_{12}\text{Expenditure2}_{ij} + \beta_{13}\text{Finanasset}_{ij},
\end{align*}
\begin{figure}
	\centering
	\includegraphics[width=1\linewidth]{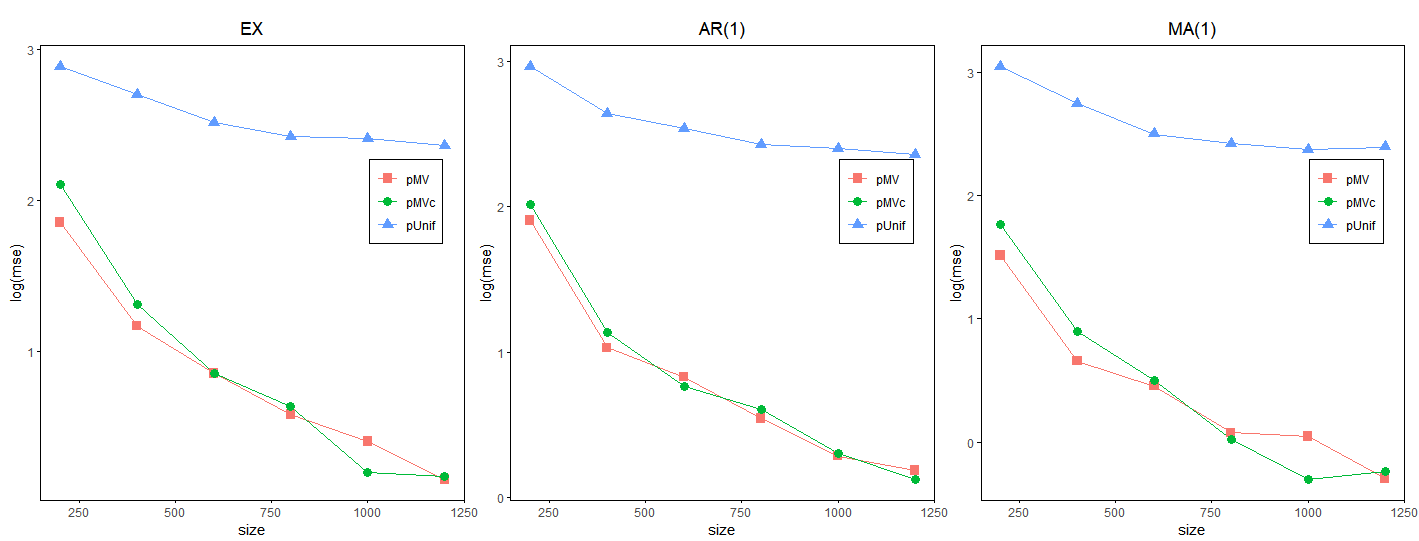}
	\caption{Log(MSE) of the estimator for the three different working correlation matrixs under the real dataset, where pUnif denotes uniform Poisson subsampling and pMV and pMVc denote optimal Poisson subsampling based on the A- and L-criteria, respectively.}
\end{figure}

Given that the true values of model parameters are typically unobservable in real-world datasets, this paper utilizes the parameter estimation values derived from the full dataset to substitute for the unknown true values. Figure 5 presents the estimation results of the pUnif and the pMV and pMVc under different working correlation matrices. The results indicate that pMV and pMVc consistently outperform pUnif. Moreover, pMV and pMVc exhibit similar performance. Therefore, it can be concluded that the optimal Poisson subsampling is a more effective choice.

\section*{Acknowledgementss}
Chunjing Li was partly supported by the National Social Science Fund of China (24BTJ061) and Scientific Research Project of Jilin Provincial Department of Education (JJKH20250702KJ).  
Xiaohui Yuan was partly supported by the National Social Science Fund of China (22BTJ019) and Scientific Research Project of Jilin Provincial Department of Science and Technology (20250102029JC).

\makeatletter
\renewenvironment{thebibliography}[1]
{\section*{\refname}%
\@mkboth{\MakeUppercase\refname}{\MakeUppercase\refname}%
\list{\@biblabel{\@arabic\c@enumiv}}%
{\settowidth\labelwidth{\@biblabel{#1}}%
\leftmargin\labelwidth \advance\leftmargin\labelsep
\advance\leftmargin by 2em%
\itemindent -2em%
\@openbib@code
\usecounter{enumiv}%
\let\p@enumiv\@empty
\renewcommand\theenumiv{\@arabic\c@enumiv}}%
\sloppy \clubpenalty4000 \@clubpenalty \clubpenalty
\widowpenalty4000%
\sfcode`\.\@m} {\def\@noitemerr
{\@latex@warning{Empty `thebibliography' environment}}%
\endlist}
\renewcommand\@biblabel[1]{}
\makeatother
\appendix
\section*{Appendix}
\newcounter{lemm}
\newtheorem{lemm}{Lemma A.}
\setcounter {equation}{0}
\def\theequation{A.\arabic{equation}}
In the context of the subsample, the weighted generalized estimating equation (\ref{2}) can be reformulated as:
\begin{equation*}
\mathbf{S}_r\left(\boldsymbol{\beta}\right) = \frac{1}{n}\sum_{i=1}^{n}\frac{\delta_i}{\pi_i}\mathbf{X}_i^{T}\mathbf{A}_i^{1/2}\left(\boldsymbol{\beta}\right)\tilde{\mathbf{R}}^{-1}\boldsymbol{\varepsilon}_{i}(\boldsymbol{\beta}) = 0.
\end{equation*}
To prove these theorems, we first establish several lemmas.

\begin{lemm}\label{lemm1}
Under assumptions (C1)-(C3), if ${p_n^2}/{r}=o(1)$ as $r\rightarrow\infty$, then
	$$\|\tilde{\mathbf{R}}^{-1}-\bar{\mathbf{R}}^{-1}\|=O_p(\sqrt{p_n/r}).$$
\end{lemm}
\noindent\textbf{\emph{The proof of Lemma \ref{lemm1}.}} We first prove that
\begin{equation}\label{1}
    \|\tilde{\mathbf{R}} - \mathbf{R}^*\| = O_p(\sqrt{p_n/r}),
\end{equation}
where $$\mathbf{R}^*=\frac{1}{n}\sum_{i=1}^n\frac{\delta_i}{\pi_i}\mathbf{A}_i^{-1/2}\left(\boldsymbol{\beta}_{0}\right)\left(\mathbf{Y}_i-\boldsymbol{\mu}_i\left(\boldsymbol{\beta}_{0}\right)\right)\left(\mathbf{Y}_i-\boldsymbol{\mu}_i\left(\boldsymbol{\beta}_{0}\right)\right)^T\mathbf{A}_i^{-1/2}\left(\boldsymbol{\beta}_{0}\right).$$
According to the Central Limit Theorem, it holds that $\|\mathbf{R}^*-\mathbf{R}_0\|=O_p(\sqrt{1/r})$. Combining with (\ref{1}), we can derive that $\|\tilde{\mathbf{R}}-\mathbf{R}_0\|=O_p(\sqrt{p_n/r})$. Note that $\|\tilde{\mathbf{R}}-\mathbf{R}^*\|^2=\sum_{k=1}^m\sum_{j=1}^m[\tilde{\mathbf{R}}_{kj}-\mathbf{R}_{kj}^*]^2$. We have
\[
\begin{aligned}
\left|\tilde{\mathbf{R}}_{kj} - \mathbf{R}_{kj}^*\right| 
& \leq \left| \frac{1}{n} \sum_{i=1}^n \frac{\delta_i}{\pi_i} \frac{(Y_{ik} - \mu_{ik}(\tilde{\boldsymbol{\beta}}))(Y_{ij} - \mu_{ij}(\tilde{\boldsymbol{\beta}})) - \left(Y_{ik} - \mu_{ik}(\boldsymbol{\beta}_{0})\right)\left(Y_{ij} - \mu_{ij}(\boldsymbol{\beta}_{0})\right) }{ \sqrt{\mathbf{A}_{ik}(\boldsymbol{\beta}_{0}) \mathbf{A}_{ij}(\boldsymbol{\beta}_{0})} } \right| \\
& \quad + \left| \frac{1}{n} \sum_{i=1}^n \frac{\delta_i}{\pi_i} \frac{ (Y_{ik} - \mu_{ik}(\tilde{\boldsymbol{\beta}}))(Y_{ij} - \mu_{ij}(\tilde{\boldsymbol{\beta}})) }{ \sqrt{\mathbf{A}_{ik}(\boldsymbol{\beta}_{0}) \mathbf{A}_{ij}(\boldsymbol{\beta}_{0})} } \eta_{ijk} \right| \\
& \triangleq I_{kj,1} + I_{kj,2},
\end{aligned}\]
where $\eta_{ijk}=\frac{\sqrt{\mathbf{A}_{ik}(\boldsymbol{\beta}_{0})\mathbf{A}_{ij}\left(\boldsymbol{\beta}_{0}\right)}}{\sqrt{\mathbf{A}_{ik}(\tilde{\boldsymbol{\beta}})\mathbf{A}_{ij}(\tilde{\boldsymbol{\beta}})}}-1$. Thus,
$$\|\tilde{\mathbf{R}}-\mathbf{R}^*\|^2\leq2\sum_{k=1}^m\sum_{j=1}^mI_{kj,1}^2+2\sum_{k=1}^m\sum_{j=1}^mI_{kj,2}^2\triangleq I_{n1}+I_{n2}.$$
By the triangle inequality, we observe
$$
\begin{aligned}
 & (Y_{ik}-\tilde{\mu}_{ik})(Y_{ij}-\tilde{\mu}_{ij})-(Y_{ik}-\mu_{ik})(Y_{ij}-\mu_{ij}) \\
 & =(Y_{ik}-\mu_{ik})(\mu_{ij}-\tilde{\mu}_{ij})+(\mu_{ik}-\tilde{\mu}_{ik})(Y_{ij}-\mu_{ij})+(\mu_{ik}-\tilde{\mu}_{ik})(\mu_{ij}-\tilde{\mu}_{ij}),
\end{aligned}$$
then, applying this result yields 
$$\begin{aligned}
I_{kj,1} & \leq\left|\frac{1}{n}\sum_{i=1}^n\frac{\delta_i}{\pi_i}\frac{(\mu_{ik}\left(\boldsymbol{\beta}_{0}\right)-\mu_{ik}(\tilde{\boldsymbol{\beta}}))(\mu_{ij}\left(\boldsymbol{\beta}_{0}\right)-\mu_{ij}(\tilde{\boldsymbol{\beta}}))}{\sqrt{\mathbf{A}_{ik}\left(\boldsymbol{\beta}_{0}\right)\mathbf{A}_{ij}\left(\boldsymbol{\beta}_{0}\right)}}\right| \\
 & \quad +\left|\frac{1}{n}\sum_{i=1}^n\frac{\delta_i}{\pi_i}\frac{(\mu_{ik}(\boldsymbol{\beta}_{0})-\mu_{ik}(\tilde{\boldsymbol{\beta}}))\left(Y_{ij}-\mu_{ij}\left(\boldsymbol{\beta}_{0}\right)\right)}{\sqrt{\mathbf{A}_{ik}\left(\boldsymbol{\beta}_{0}\right)\mathbf{A}_{ij}\left(\boldsymbol{\beta}_{0}\right)}}\right| \\
 & \quad +\left|\frac{1}{n}\sum_{i=1}^n\frac{\delta_i}{\pi_i}\frac{\left(Y_{ik}-\mu_{ik}\left(\boldsymbol{\beta}_{0}\right)\right)(\mu_{ij}\left(\boldsymbol{\beta}_{0}\right)-\mu_{ij}(\tilde{\boldsymbol{\beta}}))}{\sqrt{\mathbf{A}_{ik}\left(\boldsymbol{\beta}_{0}\right)\mathbf{A}_{ij}\left(\boldsymbol{\beta}_{0}\right)}}\right| \\
 & \triangleq I_{kj,11}+I_{kj,12}+I_{kj,13}.
\end{aligned}$$
Thus, we obtain the following result:
$$I_{n1}\leq6\sum_{k=1}^m\sum_{j=1}^mI_{kj,11}^2+6\sum_{k=1}^m\sum_{j=1}^mI_{kj,12}^2+6\sum_{k=1}^m\sum_{j=1}^mI_{kj,13}^2\triangleq I_{n11}+I_{n12}+I_{n13}.$$
By the Cauchy-Schwarz inequality,
$$ I_{kj,11}^{2} \leq \left(\frac{1}{n}\sum_{i=1}^{n}\frac{\delta_{i}}{\pi_{i}}\frac{(\mu_{ik}(\boldsymbol{\beta}_{0})-\mu_{ik}(\tilde{\boldsymbol{\beta}}))^{2}}{\mathbf{A}_{ik}(\beta_{0})}\right)\left(\frac{1}{n}\sum_{i=1}^{n}\frac{\delta_{i}}{\pi_{i}}\frac{(\mu_{ij}(\boldsymbol{\beta}_{0})-\mu_{ij}(\tilde{\boldsymbol{\beta}}))^{2}}{\mathbf{A}_{ij}(\boldsymbol{\beta}_{0})}\right). $$
Since $(\mu_{ik}(\boldsymbol{\beta}_{0})-\mu_{ik}(\tilde{\boldsymbol{\beta}}))=\mathbf{A}_{ik}(\boldsymbol{\beta}^*)\mathbf{X}_{ik}(\boldsymbol{\beta}_{0}-\tilde{\boldsymbol{\beta}}),$ $\left\|\boldsymbol{\beta}_{0}-\boldsymbol{\beta}^*\right\|\leq\|\boldsymbol{\beta}_{0}-\tilde{\boldsymbol{\beta}}\|$.
Therefore,
$$\begin{aligned}
\frac{1}{n}\sum_{i=1}^n\frac{\delta_i}{\pi_i}\frac{(\mu_{ik}(\boldsymbol{\beta}_{0})-\mu_{ik}(\tilde{\boldsymbol{\beta}}))^2}{\mathbf{A}_{ik}\left(\boldsymbol{\beta}_{0}\right)} & =\frac{1}{n}\sum_{i=1}^n\frac{\delta_i}{\pi_i}\frac{(\tilde{\boldsymbol{\beta}}-\boldsymbol{\beta}_{0})^T\mathbf{X}_{ik}\mathbf{X}_{ik}^T\mathbf{A}_{ik}^2\left(\boldsymbol{\beta}^*\right)(\tilde{\boldsymbol{\beta}}-\boldsymbol{\beta}_{0})}{\mathbf{A}_{ik}(\boldsymbol{\beta}_{0})} \\
 & \leq C(\tilde{\boldsymbol{\beta}}-\boldsymbol{\beta}_{0})^T\left(\frac{1}{n}\sum_{i=1}^n\frac{\delta_i}{\pi_i}\mathbf{X}_{ik}\mathbf{X}_{ik}^T\right)(\tilde{\boldsymbol{\beta}}-\boldsymbol{\beta}_{0}) \\
 & =O_p(\|\tilde{\boldsymbol{\beta}}-\boldsymbol{\beta}_{0}\|^2).
\end{aligned}$$
Let $C$ denote a generic positive constant. We use this notation consistently throughout the paper. Hence $I_{n11}=O_p(\|\tilde{\boldsymbol{\beta}}-\boldsymbol{\beta}_{0}\|^4)$.
Similarly, note that 
$$\begin{aligned}
I_{kj,12}^{2} & \leq\left(\frac{1}{n}\sum_{i=1}^n\frac{\delta_i}{\pi_i}\frac{(\mu_{ik}\left(\boldsymbol{\beta}_{0}\right)-\mu_{ik}(\tilde{\boldsymbol{\beta}}))^2}{\mathbf{A}_{ik}\left(\boldsymbol{\beta}_{0}\right)}\right)\left(\frac{1}{n}\sum_{i=1}^n\frac{\delta_i}{\pi_i}\frac{\left(Y_{ij}-\mu_{ij}\left(\boldsymbol{\beta}_{0}\right)\right)^2}{\mathbf{A}_{ij}\left(\boldsymbol{\beta}_{0}\right)}\right) \\
 & =O_p(\|\boldsymbol{\beta}_{0}-\tilde{\boldsymbol{\beta}}\|^2).
\end{aligned}$$
Therefore, we have $I_{n12}=O_{p}(\|\tilde{\boldsymbol{\beta}}-\boldsymbol{\beta}_{0}\|^{2})$. Similarly, $I_{n13}=O_{p}(\|\tilde{\boldsymbol{\beta}}-\boldsymbol{\beta}_{0}\|^{2})$. We next analyze $I_{kj,2}$. By applying the Cauchy–Schwarz inequality, we have
$$\begin{aligned} I_{n2} & \leq2\sum_{k=1}^m\sum_{j=1}^m\left[\frac{1}{n}\sum_{i=1}^n\frac{\delta_i}{\pi_i}\frac{(Y_{ik}-\mu_{ik}(\tilde{\boldsymbol{\beta}}))^2(Y_{ij}-\mu_{ij}(\tilde{\boldsymbol{\beta}}))^2}{\mathbf{A}_{ik}\left(\boldsymbol{\beta}_{0}\right)\mathbf{A}_{ij}\left(\boldsymbol{\beta}_{0}\right)}\right]\left[\frac{1}{n}\sum_{i=1}^n\frac{\delta_i}{\pi_i}\eta_{ijk}^2\right] \\ & \leq\frac{C}{n}\sum_{i=1}^n\sum_{k=1}^m\sum_{j=1}^m\frac{\delta_i}{n\pi_i}\eta_{ijk}^2. \end{aligned}$$
Let $g(\boldsymbol{\beta})=\frac{\sqrt{\mathbf{A}_{ik}(\boldsymbol{\beta}_{0})\mathbf{A}_{ij}(\boldsymbol{\beta}_{0})}}{\sqrt{\mathbf{A}_{ik}(\boldsymbol{\beta})\mathbf{A}_{ij}(\boldsymbol{\beta})}}$, we have, $\eta_{ijk}=g(\tilde{\boldsymbol{\beta}})-g\left(\boldsymbol{\beta}_{0}\right)$, then
$$\begin{aligned}
\frac{\partial g(\boldsymbol{\beta})}{\partial\boldsymbol{\beta}} &= \sqrt{\mathbf{A}_{ik}(\boldsymbol{\beta}_{0})\mathbf{A}_{ij}(\boldsymbol{\beta}_{0})} \left[
   -\frac{1}{2}\mathbf{A}_{ij}^{-3/2}(\boldsymbol{\beta})\dot{\mathbf{A}}_{ij}(\boldsymbol{\beta})\mathbf{X}_{ij}\mathbf{A}_{ik}^{-1/2}(\boldsymbol{\beta}) \right. \\
&\quad \left. -\frac{1}{2}\mathbf{A}_{ij}^{-1/2}(\boldsymbol{\beta})\mathbf{A}_{ik}^{-3/2}(\boldsymbol{\beta})\dot{\mathbf{A}}_{ik}(\boldsymbol{\beta})\mathbf{X}_{ik}
\right]\\
 & =-\frac{1}{2}g(\boldsymbol{\beta})\left[\frac{\dot{\mathbf{A}}_{ij}(\boldsymbol{\beta})}{\mathbf{A}_{ij}(\boldsymbol{\beta})}\mathbf{X}_{ij}+\frac{\dot{\mathbf{A}}_{ik}(\boldsymbol{\beta})}{\mathbf{A}_{ik}(\boldsymbol{\beta})}\mathbf{X}_{ik}\right].
\end{aligned}$$
Therefore,
$$\begin{aligned}
I_{n2} & \leq\frac{C}{n}\sum_{i=1}^{n}\sum_{k=1}^{m}\sum_{j=1}^{m}\frac{\delta_{i}}{\pi_{i}}(\tilde{\boldsymbol{\beta}}-\boldsymbol{\beta}_{0})^{T}\left[ g(\boldsymbol{\beta}^{*})\left(\frac{\dot{\mathbf{A}}_{ij}(\boldsymbol{\beta}^{*})}{\mathbf{A}_{ij}(\boldsymbol{\beta}^{*})}\mathbf{X}_{ij}+\frac{\dot{\mathbf{A}}_{ik}(\boldsymbol{\beta}^{*})}{\mathbf{A}_{ik}(\boldsymbol{\beta}^{*})}X_{ik}\right)\right] \\
 & \quad \left[g(\boldsymbol{\beta}^*)\left(\frac{\dot{\mathbf{A}}_{ij}(\boldsymbol{\beta}^*)}{\mathbf{A}_{ij}(\boldsymbol{\beta}^*)}\mathbf{X}_{ij}+\frac{\dot{\mathbf{A}}_{ik}(\boldsymbol{\beta}^*)}{\mathbf{A}_{ik}(\boldsymbol{\beta}^*)}\mathbf{X}_{ik}\right)\right]^T(\tilde{\boldsymbol{\beta}}-\boldsymbol{\beta}_{0})\\
 & \leq\frac{2C}{n}\sum_{i=1}^n\sum_{k=1}^m\sum_{j=1}^m\frac{\delta_i}{\pi_i}(\tilde{\boldsymbol{\beta}}-\boldsymbol{\beta}_{0})^T\left(g^2(\boldsymbol{\beta}^*)\mathbf{X}_{ij}\mathbf{X}_{ij}^T+g^2(\boldsymbol{\beta}^*)\mathbf{X}_{ik}\mathbf{X}_{ik}^T\right)(\tilde{\boldsymbol{\beta}}-\boldsymbol{\beta}_{0})\\
 & \leq C\|\tilde{\boldsymbol{\beta}}-\boldsymbol{\beta}_{0}\|^2\left[\lambda_{\max}\left(\frac{1}{n}\sum_{i=1}^n\frac{\delta_i}{\pi_i}\mathbf{X}_i^T\mathbf{X}_i\right)+\lambda_{\max}\left(\frac{1}{n}\sum_{i=1}^n\frac{\delta_i}{\pi_i}\mathbf{X}_i^T\mathbf{X}_i\right)\right] \\
 & =O_p(\|\tilde{\boldsymbol{\beta}}-\boldsymbol{\beta}_{0}\|^2), 
\end{aligned}$$
where $\left\|\boldsymbol{\beta}_{0}-\boldsymbol{\beta}^*\right\|\leq\|\boldsymbol{\beta}_{0}-\tilde{\boldsymbol{\beta}}\|$.
Hence, under Assumption (C3) and Lemma \ref{lemm3}, $\|\tilde{\mathbf{R}}-\mathbf{R}^*\|^2\leq I_{n1}+I_{n2}=O_p(\|\tilde{\boldsymbol{\beta}}-\boldsymbol{\beta}_{0}\|^2)=O_p(p_n/r)$, that is, (\ref{1}) is proved. Since 
$\|\tilde{\mathbf{R}}-\mathbf{R}_0\|\leq\|\tilde{\mathbf{R}}-\mathbf{R}^*\|+\|\mathbf{R}^*-\mathbf{R}_0\|=O_p(\sqrt{p_n/r})$. Therefore, we have $\|\tilde{\mathbf{R}}^{-1}-\mathbf{R}_0^{-1}\|=\|\tilde{\mathbf{R}}^{-1}(\tilde{\mathbf{R}}-\mathbf{R}_0)\mathbf{R}_0^{-1}\|=O_p(\sqrt{p_n/r})$.


\begin{lemm}\label{lemm2}
	Under assumptions (C1)-(C7), if  $p_n^2/r=o(1)$, then $\|\mathbf{S}_r(\boldsymbol{\beta}_{0})-\bar{\mathbf{S}}_r(\boldsymbol{\beta}_{0})\|=O_p\left(p_n/r\right).$
\end{lemm}
\noindent\textbf{\emph{The proof of Lemma \ref{lemm2}.}} Let $\mathbf{Q}=\left\{q_{j_1j_2}\right\}_{1\leq j_1,j_2\leq m}$ denote $\tilde{\mathbf{R}}^{-1}-\bar{\mathbf{R}}^{-1}$. Then 
\begin{align*}
\mathbf{S}_r(\boldsymbol{\beta}_{0}) - \bar{\mathbf{S}}_r(\boldsymbol{\beta}_{0}) 
&= \frac{1}{n}\sum_{i=1}^n \frac{\delta_i}{\pi_i} \mathbf{X}_i^T \mathbf{A}_i^{1/2}(\boldsymbol{\beta}_{0}) \mathbf{Q} \mathbf{A}_i^{-1/2}(\boldsymbol{\beta}_{0}) (\mathbf{Y}_i - \boldsymbol{\mu}_i(\boldsymbol{\beta}_{0})) \\
&= \sum_{j_1=1}^m \sum_{j_2=1}^m q_{j_1j_2} \left[ \frac{1}{n} \sum_{i=1}^n \frac{\delta_i}{\pi_i} \mathbf{A}_{ij_1}^{1/2}(\boldsymbol{\beta}_{0}) \varepsilon_{ij_2}(\boldsymbol{\beta}_{0}) \mathbf{X}_{ij_1} \right],
\end{align*}
where $\varepsilon_{ij_2}(\boldsymbol{\beta}_{0})=\mathbf{A}_{ij_2}^{-1/2}(\boldsymbol{\beta}_{0})(Y_{ij_2}-\mu_{ij_2}(\boldsymbol{\beta}_{0}))$. Since
$$\begin{aligned}
E\left[\left\|\frac{1}{n}\sum_{i=1}^{n}\frac{\delta_{i}}{\pi_{i}}\mathbf{A}_{ij_{1}}^{1/2}\left(\boldsymbol{\beta}_{0}\right)\varepsilon_{ij_{2}}\left(\boldsymbol{\beta}_{0}\right)\mathbf{X}_{ij_{1}}\right\|^{2}\right] & =\frac{1}{n^2}\sum_{i=1}^n\frac{1}{\pi_i}\mathbf{A}_{ij_1}\left(\boldsymbol{\beta}_{0}\right)E\left[\boldsymbol{\varepsilon}_{ij_2}^2\left(\boldsymbol{\beta}_{0}\right)\right]\mathbf{X}_{ij_1}^T\mathbf{X}_{ij_1} \\
 & \leq\frac{1}{n^2}\sum_{i=1}^n\frac{1}{\pi_i}\mathbf{X}_{ij_1}^T\mathbf{X}_{ij_1} \\
 & \leq\left(\max_{1\leq i\leq n}\frac{1}{n\pi_i}\right)\sum_{i=1}^n\frac{1}{n}\mathbf{X}_{ij_1}^T\mathbf{X}_{ij_1} \\
 & =O_p(p_{n}/r).
\end{aligned}$$
Thus, we have 
$\|\frac{1}{n}\sum_{i=1}^n\frac{\delta_i}{\pi_i}\mathbf{A}_{ij_1}^{1/2}(\boldsymbol{\beta}_{0})\varepsilon_{ij_2}(\boldsymbol{\beta}_{0})\mathbf{X}_{ij_1}\|=O_p(\sqrt{p_{n}/r}), \forall1\leq j_1,j_2\leq m.$ Combining with assumptions (C4) yields $q_{j_1j_2}=O_p(\sqrt{p_n/r}),\forall1\leq j_1,j_2\leq m,$ which completes the proof.

\begin{lemm}\label{lemm3} if ${p_n^2}/r=o(1)$, 
$\tilde{\mathbf{S}}_r(\boldsymbol{\beta}) = \frac{1}{n}\sum_{i=1}^n \frac{\delta_i}{\pi_i} \mathbf{X}_i^T (\mathbf{Y}_i - \boldsymbol{\mu}_i(\boldsymbol{\beta})),$
where $\tilde{\boldsymbol{\beta}}$ is the solution to $\tilde{\mathbf{S}}_r(\boldsymbol{\beta}) = 0$, then $
\|\tilde{\boldsymbol{\beta}} - \boldsymbol{\beta}_{0}\| = O_p(\sqrt{p_n/r}).$
\end{lemm}
\noindent\textbf{\emph{The proof of Lemma \ref{lemm3}.}} It suffices to show that for any $\epsilon > 0$, there exists $\Delta>0$ such that for sufficiently large $r$, we have 
$$P\left(\sup_{\|\boldsymbol{\beta}-\boldsymbol{\beta}_{0}\|=\Delta\sqrt{p_{n}/r}}\left(\boldsymbol{\beta}-\boldsymbol{\beta}_{0}\right)^T\tilde{\mathbf{S}}_r(\boldsymbol{\beta})<0\right)\geq1-\epsilon.$$
Since
$$
\begin{aligned}
(\boldsymbol{\beta} - \boldsymbol{\beta}_{0})^{T} \tilde{\mathbf{S}}_r(\boldsymbol{\beta}) &= (\boldsymbol{\beta} - \boldsymbol{\beta}_{0})^T \tilde{\mathbf{S}}_r(\boldsymbol{\beta}_{0}) + (\boldsymbol{\beta} - \boldsymbol{\beta}_{0})^T \frac{\partial \tilde{\mathbf{S}}_r(\boldsymbol{\beta}^*)}{\partial \boldsymbol{\beta}^T} (\boldsymbol{\beta} - \boldsymbol{\beta}_{0}) \\
&\triangleq I_{n1} + I_{n2},
\end{aligned}
$$
where $\|\boldsymbol{\beta}^*-\boldsymbol{\beta}_{0}\|\leq\|\boldsymbol{\beta}-\boldsymbol{\beta}_{0}\|.$ We now analyse the term $I_{n1}.$ Applying the Cauchy–Schwarz inequality yields $|I_{n1}|\leq\|\boldsymbol{\beta}-\boldsymbol{\beta}_{0}\|\|\tilde{\mathbf{S}}_r(\boldsymbol{\beta}_{0})\|=\Delta\sqrt{{p_n}/{r}}\|\tilde{\mathbf{S}}_r(\boldsymbol{\beta}_{0})\|.$ Note that 
$$\begin{aligned}
E\left[\|\tilde{\mathbf{S}}_r\left(\boldsymbol{\beta}_{0}\right)\|^{2}\right] & =E\left[\frac{1}{n^2}\sum_{i=1}^n\frac{1}{\pi_i}\left(\mathbf{Y}_i-\boldsymbol{\mu}_i\left(\boldsymbol{\beta}_{0}\right)\right)^T\mathbf{X}_i\mathbf{X}_i^T\left(\mathbf{Y}_i-\boldsymbol{\mu}_i\left(\boldsymbol{\beta}_{0}\right)\right)\right] \\
 & \leq C\left(\max_{1\leq i\leq n}\frac{1}{n\pi_i}\right)\lambda_{\min}\left(\frac{1}{n}\sum_{i=1}^n\mathbf{X}_i\mathbf{X}_i^T\right) \\
 & \leq C\left(\max_{1\leq i\leq n}\frac{1}{n\pi_i}\right)Tr\left(\frac{1}{n}\sum_{i=1}^n\sum_{j=1}^m\mathbf{X}_{ij}^T\mathbf{X}_{ij}\right) \\
 & =O_p\left({p_n}/{r}\right).
\end{aligned}$$
Therefore, $\left|I_{n1}\right|\leq\Delta\sqrt{{p_{n}}/{r}}O_{p}(\sqrt{{p_{n}}/{r}})=\Delta O_{p}\left({p_{n}}/{r}\right).$ Next, we consider $I_{n2},$ 
$$\begin{aligned}
I_{n2} & =-\left(\boldsymbol{\beta}-\boldsymbol{\beta}_{0}\right)^T\left[\frac{1}{n}\sum_{i=1}^n\frac{\delta_i}{\pi_i}\mathbf{X}_i^T\mathbf{A}_i\left(\boldsymbol{\beta}^*\right)\mathbf{X}_i\right]\left(\boldsymbol{\beta}-\boldsymbol{\beta}_{0}\right) \\
 & =-\left(\boldsymbol{\beta}-\boldsymbol{\beta}_{0}\right)^T\left[\frac{1}{n}\sum_{i=1}^n\frac{\delta_i}{\pi_i}\mathbf{X}_i^T\mathbf{A}_i\left(\boldsymbol{\beta}_{0}\right)\mathbf{X}_i\right]\left(\boldsymbol{\beta}-\boldsymbol{\beta}_{0}\right) \\
  & \quad -\left(\boldsymbol{\beta}-\boldsymbol{\beta}_{0}\right)^T\left[\frac{1}{n}\sum_{i=1}^n\frac{\delta_i}{\pi_i}\mathbf{X}_i^T\left(\mathbf{A}_i\left(\boldsymbol{\beta}^*\right)-\mathbf{A}_i\left(\boldsymbol{\beta}_{0}\right)\right)\mathbf{X}_i\right]\left(\boldsymbol{\beta}-\boldsymbol{\beta}_{0}\right) \\
 & =I_{n21}+I_{n22}.
\end{aligned}$$
where $\frac{\partial\tilde{\mathbf{S}}\left(\boldsymbol{\beta}^*\right)}{\partial\boldsymbol{\beta}}=-\frac{1}{n}\sum_{i=1}^n\frac{\delta_i}{\pi_i}\mathbf{X}_i^T\mathbf{A}_i\left(\boldsymbol{\beta}^*\right)\mathbf{X}_i.$ We have 
$$\begin{aligned}
I_{n21} & \leq-\lambda_{\min}\left(\mathbf{A}_i\left(\boldsymbol{\beta}_{0}\right)\right)\lambda_{\min}\left(\frac{1}{n}\sum_{i=1}^n\frac{\delta_i}{\pi_i}\mathbf{X}_i^T\mathbf{X}_i\right)\left\|\boldsymbol{\beta}-\boldsymbol{\beta}_{0}\right\|^2 \\
 & =-\Delta^2O_p\left({p_n}/{r}\right),
\end{aligned}$$
and
$$\begin{aligned} \begin{vmatrix} I_{n22} \end{vmatrix} & =\left|\left(\boldsymbol{\beta}-\boldsymbol{\beta}_{0}\right)^T\left[\frac{1}{n}\sum_{i=1}^n\frac{\delta_i}{\pi_i}\mathbf{X}_i^T\left(\mathbf{A}_i\left(\boldsymbol{\beta}^*\right)-\mathbf{A}_i\left(\boldsymbol{\beta}_{0}\right)\right)\mathbf{X}_i\right]\left(\boldsymbol{\beta}-\boldsymbol{\beta}_{0}\right)\right| \\ & \leq\sup_i\left\|\mathbf{X}_i\right\|\left\|\boldsymbol{\beta}^*-\boldsymbol{\beta}_{0}\right\|\lambda_{\max}\left(\frac{1}{n}\sum_{i=1}^n\frac{\delta_i}{\pi_i}\mathbf{X}_i^T\mathbf{X}_i\right)\left\|\boldsymbol{\beta}-\boldsymbol{\beta}_{0}\right\|^2 \\ & =O_p\left(\sqrt{p_n}\right)O_p\left(\sqrt{{p_n}/{r}}\right)\Delta^2{p_n}/{r} \\ & =\Delta^2o_p\left({p_n}/{r}\right), \end{aligned}$$
where $\mathbf{A}_i\left(\boldsymbol{\beta}^*\right)-\mathbf{A}_i\left(\boldsymbol{\beta}_{0}\right)=\mathbf{A}_i\left(\boldsymbol{\beta}^{**}\right)\mathbf{X}_i\left(\boldsymbol{\beta}^*-\boldsymbol{\beta}_{0}\right),\left\|\boldsymbol{\beta}^{**}-\boldsymbol{\beta}_{0}\right\|\leq\left\|\boldsymbol{\beta}^*-\boldsymbol{\beta}_{0}\right\|.$ When $\Delta$ is sufficiently large, $\left(\boldsymbol{\beta}-\boldsymbol{\beta}_{0}\right)^T\tilde{\mathbf{S}}_r\left(\boldsymbol{\beta}\right)$ is dominated by $I_{n21}$, and for sufficiently large $r$, this value can be made negative.

\begin{lemm}\label{lemm4} Let $\bar{\mathbf{D}}_r\left(\boldsymbol{\beta}\right)=-\frac{\partial \bar{\mathbf{S}}_r(\boldsymbol{\beta})}{\partial \boldsymbol{\beta}^T}$, we have
$$\bar{\mathbf{D}}_r\left(\boldsymbol{\beta}\right)=\bar{\mathbf{H}}_r\left(\boldsymbol{\beta}\right)+\bar{\mathbf{E}}_r\left(\boldsymbol{\beta}\right)+\bar{\mathbf{G}}_r\left(\boldsymbol{\beta}\right),$$
where
$$\begin{aligned}
 & \bar{\mathbf{H}}_r\left(\boldsymbol{\beta}\right)=\frac{1}{n}\sum_{i=1}^n\frac{\delta_i}{\pi_i}\mathbf{X}_i^T\mathbf{A}_i^{1/2}\left(\boldsymbol{\beta}\right)\bar{\mathbf{R}}^{-1}\mathbf{A}_i^{1/2}\left(\boldsymbol{\beta}\right)\mathbf{X}_i, \\
 & \bar{\mathbf{E}}_r\left(\boldsymbol{\beta}\right)=\frac{1}{2n}\sum_{i=1}^n\frac{\delta_i}{\pi_i}\mathbf{X}_i^T\mathbf{A}_i^{1/2}\left(\boldsymbol{\beta}\right)\bar{\mathbf{R}}^{-1}\mathbf{A}_i^{-3/2}(\boldsymbol{\beta})\mathbf{C}_i\left(\boldsymbol{\beta}\right)\mathbf{F}_i\left(\boldsymbol{\beta}\right)\mathbf{X}_i, \\
 & \bar{\mathbf{G}}_r\left(\boldsymbol{\beta}\right)=-\frac{1}{2n}\sum_{i=1}^n\frac{\delta_i}{\pi_i}\mathbf{X}_i^T\mathbf{A}_i^{1/2}\left(\boldsymbol{\beta}\right)\mathbf{F}_i\left(\boldsymbol{\beta}\right)\mathbf{J}_i\left(\boldsymbol{\beta}\right)\mathbf{X}_i,
\end{aligned}$$
with
$$\begin{aligned}
 & \mathbf{C}_i\left(\boldsymbol{\beta}\right)=diag\left(Y_{i1}-\mu_{i1}\left(\boldsymbol{\beta}\right),\cdots,Y_{im}-\mu_{im}\left(\boldsymbol{\beta}\right)\right), \\
 & \mathbf{F}_i\left(\boldsymbol{\beta}\right)=diag(\ddot{\mu}(\mathbf{X}_{i1}^T\boldsymbol{\beta}),\cdots,\ddot{\mu}(\mathbf{X}_{im}^T\boldsymbol{\beta})), \\
 & \mathbf{J}_i\left(\boldsymbol{\beta}\right)=diag(\bar{\mathbf{R}}^{-1}\mathbf{A}_i^{-1/2}\left(\boldsymbol{\beta}\right)\left(\mathbf{Y}_i-\mu_i\left(\boldsymbol{\beta}\right)\right)).
\end{aligned}$$
\end{lemm}
\noindent\textbf{\emph{The proof of Lemma \ref{lemm4}.}}
See Xie \& Wang (2003).

\noindent\begin{lemm}\label{lemm5}
 Under assumptions (C1)-(C7), if ${p_n^2}/r=o(1)$, then for any $\Delta > 0$ and $\mathbf{b}_n \in \mathbf{R}^{p_n}$, we have
$$\sup_{\left\|\boldsymbol{\beta}-\boldsymbol{\beta}_{0}\right\|\leq\Delta\sqrt{{p_n}/{r}}}\sup_{\left\|\mathbf{b}_n\right\|=1}\left|\mathbf{b}_n^T\left[\mathbf{D}_r\left(\boldsymbol{\beta}\right)-\bar{\mathbf{D}}_r\left(\boldsymbol{\beta}\right)\right]\mathbf{b}_n\right|=O_p(\sqrt{{p_n}/{r}}).$$
The matrix $\mathbf{D}_r\left(\boldsymbol{\beta}\right)-\bar{\mathbf{D}}_r\left(\boldsymbol{\beta}\right)$ is symmetric, and these results follow directly,
$$\begin{gathered}
\sup_{\left\|\boldsymbol{\beta}-\boldsymbol{\beta}_{0}\right\|\leq\Delta\sqrt{{p_n}/{r}}}\left|\lambda_{\min}\left[\mathbf{D}_r\left(\boldsymbol{\beta}\right)-\bar{\mathbf{D}}_r\left(\boldsymbol{\beta}\right)\right]\right|=O_p(\sqrt{{p_n}/{r}}), \\
\sup_{\|\boldsymbol{\beta}-\boldsymbol{\beta}_{0}\|\leq\Delta\sqrt{{p_n}/{r}}}\left|\lambda_{\max}\left[\mathbf{D}_r\left(\boldsymbol{\beta}\right)-\bar{\mathbf{D}}_r\left(\boldsymbol{\beta}\right)\right]\right|=O_p(\sqrt{{p_n}/{r}}).
\end{gathered}$$
\end{lemm}

\noindent\textbf{\emph{The proof of Lemma \ref{lemm5}.}} 
 Let $\mathbf{H}_r(\boldsymbol{\beta})$,$\mathbf{E}_r(\boldsymbol{\beta})$,$\mathbf{G}_r(\boldsymbol{\beta})$ be defined analogously to $\bar{\mathbf{H}}_r\left(\boldsymbol{\beta}\right)$, $\bar{\mathbf{E}}_r\left(\boldsymbol{\beta}\right)$, $\bar{\mathbf{G}}_r\left(\boldsymbol{\beta}\right)$, but with $\tilde{\mathbf{R}}$ replacing $\bar{\mathbf{R}}$. The proof can be completed by establishing the following three asymptotic results:
 \begin{align}
&\sup_{\left\|\boldsymbol{\beta}-\boldsymbol{\beta}_{0}\right\|\leq\Delta\sqrt{{p_n}/{r}}}\sup_{\left\|\mathbf{b}_n\right\|=1}\left|\mathbf{b}_n^T\left[\mathbf{H}_r\left(\boldsymbol{\beta}\right)-\bar{\mathbf{H}}_r\left(\boldsymbol{\beta}\right)\right]\mathbf{b}_n\right|=O_p(\sqrt{{p_n}/{r}}), \label{eq1} \\
&\sup_{\left\|\boldsymbol{\beta}-\boldsymbol{\beta}_{0}\right\|\leq\Delta\sqrt{{p_n}/{r}}}\sup_{\left\|\mathbf{b}_n\right\|=1}\left|\mathbf{b}_n^T\left[\mathbf{E}_r\left(\boldsymbol{\beta}\right)-\bar{\mathbf{E}}_r\left(\boldsymbol{\beta}\right)\right]\mathbf{b}_n\right|=O_p(\sqrt{{p_n}/{r}}), \label{eq2} \\
&\sup_{\left\|\boldsymbol{\beta}-\boldsymbol{\beta}_{0}\right\|\leq\Delta\sqrt{{p_n}/{r}}}\sup_{\left\|\mathbf{b}_n\right\|=1}\left|\mathbf{b}_n^T\left[\mathbf{G}_r\left(\boldsymbol{\beta}\right)-\bar{\mathbf{G}}_r\left(\boldsymbol{\beta}\right)\right]\mathbf{b}_n\right|=O_p(\sqrt{{p_n}/{r}}). \label{eq3}
\end{align}
For (\ref{eq1}), we observe that 
$$\begin{aligned}
\left|\mathbf{b}_n^T\left[\mathbf{H}_r\left(\boldsymbol{\beta}\right)-\bar{\mathbf{H}}_r\left(\boldsymbol{\beta}\right)\right]\mathbf{b}_n\right| & \leq\|\tilde{\mathbf{R}}^{-1}-\bar{\mathbf{R}}^{-1}\|\lambda_{\max}\left(\mathbf{A}_i\left(\boldsymbol{\beta}\right)\right)\lambda_{\max}\left(\frac{1}{n}\sum_{i=1}^n\frac{\delta_i}{\pi_i}\mathbf{X}_i^T\mathbf{X}_i\right) \\
 & =O_p(\sqrt{{p_n}/{r}}).
\end{aligned}$$
By assumptions (C2) and (C4), (\ref{eq1}) is thus proved. Similarly, (\ref{eq2}) and (\ref{eq3}) can be verified.


\noindent\begin{lemm}\label{lemm6}
 Under assumptions (C1)-(C7), if ${p_n^2}/r=o(1)$, then for any $\Delta > 0$ and $\mathbf{b}_n \in \mathbf{R}^{p_n}$, we have
\begin{align*}
\sup_{\|\boldsymbol{\beta} - \boldsymbol{\beta}_{0}\| \leq \Delta\sqrt{{p_n}/{r}}}\sup_{\|\mathbf{b}_n\| = 1}\left|\mathbf{b}_n^T\bar{\mathbf{E}}_r(\boldsymbol{\beta})\mathbf{b}_n\right| &= O_p\left({p_n}/{\sqrt{r}}\right), \\
\sup_{\|\boldsymbol{\beta} - \boldsymbol{\beta}_{0}\| \leq \Delta\sqrt{{p_n}/{r}}}\sup_{\|\mathbf{b}_n\| = 1}\left|\mathbf{b}_n^T\bar{\mathbf{G}}_r(\boldsymbol{\beta})\mathbf{b}_n\right| &= O_p\left({p_n}/{\sqrt{r}}\right).
\end{align*}
\end{lemm}

\noindent\textbf{\emph{The proof of Lemma \ref{lemm6}.}} 
The proof follows directly from Lemma 3.4 in Wang (2011)


\noindent\begin{lemm}\label{lemm7}
 Under assumptions (C1)-(C7), if ${p_n^3}/r=o(1)$, then for any $\mathbf{b}_n \in \mathbf{R}^{p_n}$ with $\left\|\mathbf{b}_n\right\|=1$, we have
 $$\mathbf{b}_n^T\bar{\mathbf{M}}_r^{-1/2}\left(\boldsymbol{\beta}_{0}\right)\bar{\mathbf{S}}_r\left(\boldsymbol{\beta}_{0}\right)\xrightarrow{d}N\left(0,1\right).$$
\end{lemm}

\noindent\textbf{\emph{The proof of Lemma \ref{lemm7}.}}  
Let $Var(\bar{\mathbf{S}}_r\left(\boldsymbol{\beta}_{0}\right))=\frac{1} {n^2}\sum_{i=1}^nVar\left(\mathbf{T}_i\right)$, where 
$$\mathbf{T}_i=\frac{\delta_i}{\pi_i}\mathbf{X}_i^T\mathbf{A}_i^{1/2}\left(\boldsymbol{\beta}_{0}\right)\bar{\mathbf{R}}^{-1}\mathbf{A}_i^{-1/2}\left(\boldsymbol{\beta}_{0}\right)\left(\mathbf{Y}_i-\boldsymbol{\mu}_i\left(\boldsymbol{\beta}_{0}\right)\right).$$
Note that $Var\left(\mathbf{T}_{i}\right)=Var\left(E\left(\mathbf{T}_{i}\mid \mathbf{Y}_{i}\right)\right)+E\left(Var\left(\mathbf{T}_{i}\mid \mathbf{Y}_{i}\right)\right).$ We have 
$$E\left(\mathbf{T}_i\mid \mathbf{Y}_i\right)=\mathbf{X}_i^T\mathbf{A}_i^{1/2}\left(\boldsymbol{\beta}_{0}\right)\bar{\mathbf{R}}^{-1}\boldsymbol{\varepsilon}_i\left(\boldsymbol{\beta}_{0}\right),$$
and
$$Var\left(\mathbf{T}_i\mid \mathbf{Y}_i\right)=\frac{1-\pi_i}{\pi_i}\mathbf{X}_i^T\mathbf{A}_i^{1/2}\left(\boldsymbol{\beta}_{0}\right)\bar{\mathbf{R}}^{-1}\boldsymbol{\varepsilon}_i\left(\boldsymbol{\beta}_{0}\right)\boldsymbol{\varepsilon}_i^T\left(\boldsymbol{\beta}_{0}\right)\bar{\mathbf{R}}^{-1}\mathbf{A}_i^{1/2}\left(\boldsymbol{\beta}_{0}\right)\mathbf{X}_i.$$
Thus, we obtain $\bar{\mathbf{M}}_r\left(\boldsymbol{\beta}_{0}\right)=\frac{1}{n^2}\sum_{i=1}^n\frac{1}{\pi_i}\mathbf{X}_i^T\mathbf{A}_i^{1/2}\left(\boldsymbol{\beta}_{0}\right)\bar{\mathbf{R}}^{-1}\boldsymbol{\varepsilon}_i\left(\boldsymbol{\beta}_{0}\right)\boldsymbol{\varepsilon}_i^T\left(\boldsymbol{\beta}_{0}\right)\bar{\mathbf{R}}^{-1}\mathbf{A}_i^{1/2}\left(\boldsymbol{\beta}_{0}\right)\mathbf{X}_i.$
 Let $\mathbf{b}_n^T\bar{\mathbf{M}}_r^{-1/2}\left(\boldsymbol{\beta}_{0}\right)\bar{\mathbf{S}}_r\left(\boldsymbol{\beta}_{0}\right)=\sum_{i=1}^n\mathbf{Z}_i,$ where $$\mathbf{Z}_i=\frac{\delta_i}{n\pi_i}\mathbf{b}_n^T\bar{\mathbf{M}}_r^{-1/2}\left(\boldsymbol{\beta}_{0}\right)\mathbf{X}_i^T\mathbf{A}_i^{1/2}\left(\boldsymbol{\beta}_{0}\right)\bar{\mathbf{R}}^{-1}\boldsymbol{\varepsilon}_i\left(\boldsymbol{\beta}_{0}\right).$$ Then, $E\left(\mathbf{Z}_i\right)=0,Var\left(\sum_{i=1}^n\mathbf{Z}_i\right)=\mathbf{b}_n^T\mathbf{b}_n=1.$ Note that 
$$\begin{aligned}
\mathbf{Z}_i^2 & =\left[\frac{\delta_i}{n\pi_i}\mathbf{b}_n^T\bar{\mathbf{M}}_r^{-1/2}\left(\boldsymbol{\beta}_{0}\right)\mathbf{X}_i^T\mathbf{A}_i^{1/2}\left(\boldsymbol{\beta}_{0}\right)\bar{\mathbf{R}}^{-1}\boldsymbol{\varepsilon}_i\left(\boldsymbol{\beta}_{0}\right)\right]^2 \\
 & \leq\frac{1}{n^2}\left[\mathbf{b}_n^T\bar{\mathbf{M}}_r^{-1/2}\left(\boldsymbol{\beta}_{0}\right)\mathbf{X}_i^T\mathbf{A}_i^{1/2}\left(\boldsymbol{\beta}_{0}\right)\bar{\mathbf{R}}^{-2}\mathbf{A}_i^{1/2}\left(\boldsymbol{\beta}_{0}\right)\mathbf{X}_i\bar{\mathbf{M}}_r^{-1/2}\left(\boldsymbol{\beta}_{0}\right)\mathbf{b}_n\right]\left\|\frac{\delta_i}{\pi_i}\boldsymbol{\varepsilon}_i\left(\boldsymbol{\beta}_{0}\right)\right\|^2.
\end{aligned}$$
Let 
$$\begin{aligned}
\xi_{i} & =\mathbf{b}_n^T\bar{\mathbf{M}}_r^{-1/2}\left(\boldsymbol{\beta}_{0}\right)\mathbf{X}_i^T\mathbf{A}_i^{1/2}\left(\boldsymbol{\beta}_{0}\right)\bar{\mathbf{R}}^{-2}\mathbf{A}_i^{1/2}\left(\boldsymbol{\beta}_{0}\right)\mathbf{X}_i\bar{\mathbf{M}}_r^{-1/2}\left(\boldsymbol{\beta}_{0}\right)\mathbf{b}_n \\
 & \leq\lambda_{\max}(\mathbf{X}_i^T\mathbf{X}_i)\lambda_{\min}^{-1}(\bar{\mathbf{M}}_r\left(\boldsymbol{\beta}_{0}\right))\lambda_{\max}(\bar{\mathbf{R}}^{-2})\lambda_{\max}\left(\mathbf{A}_i(\boldsymbol{\beta}_{0})\right) \\
 & =C\frac{\lambda_{\max}(\mathbf{X}_i^T\mathbf{X}_i)}{\lambda_{\min}(\bar{\mathbf{M}}_r\left(\boldsymbol{\beta}_{0}\right))}.
\end{aligned}$$
Since 
\begin{equation}\label{A5}
\begin{aligned}
\mathbf{b}_n^T \bar{\mathbf{M}}_r(\boldsymbol{\beta}_{0}) \mathbf{b}_n &\geq C \frac{1}{n} \left(\sum_{i=1}^n \frac{1}{n\pi_i} \mathbf{X}_i^T \mathbf{X}_i\right)= O_p\left(1/{r}\right).
\end{aligned}
\end{equation}
Thus, $\max_{1\leq i\leq n}\xi_i=\frac{O\left(p_n\right)}{O(1/{r})}=O_p\left({rp_n}\right). $We now verify the Lyapunov condition. 
$$\begin{aligned}
\sum_{i=1}^nE\left|\mathbf{Z}_i\right|^{2+\delta} & =\sum_{i=1}^nE\left(\left|\mathbf{Z}_i\right|^2\right)^{(2+\delta)/2} \\
 & \leq\sum_{i=1}^nE\left(\frac{1}{n^2}\xi_i\left\|\frac{\delta_i}{\pi_i}\boldsymbol{\varepsilon}_i\left(\boldsymbol{\beta}_{0}\right)\right\|^2\right)^{(2+\delta)/2} \\
 & =\sum_{i=1}^n\left(\frac{1}{n^{2+\delta}}\xi_i^{1+\delta/2}\frac{\delta_i}{\pi_i^{2+\delta}}\right)E\left(\left\|\boldsymbol{\varepsilon}_i\left(\boldsymbol{\beta}_{0}\right)\right\|^{2+\delta}\right) \\
 & \leq\frac{1}{n^{1+\delta}}\left(\max_{1\leq i\leq n}\xi_i\right)^{\delta/2}\left(\max_{1\leq i\leq n}\frac{1}{\pi_i}\right)^{1+\delta}\left(\frac{1}{n}\sum_{i=1}^n\xi_i\right) \\
 & =\frac{1}{n^{1+\delta}}O_p\left(\left({rp_n}\right)^{\delta/2}\right)O_p\left(\left(\frac{n}{r}\right)^{1+\delta}\right)\left(\frac{1}{n}\sum_{i=1}^n\xi_i\right).
\end{aligned}$$
Observe that $\frac{1}{n}\sum_{i=1}^{n}\xi_{i}\leq\frac{\lambda_{\max}\left(\frac{1}{n}\sum_{i=1}^{n}\mathbf{X}_{i}^{T}\mathbf{X}_{i}\right)}{\lambda_{\min}\left(\bar{\mathbf{M}}\left(\boldsymbol{\beta}_{0}\right)\right)}=O_p\left({r}\right).$ Thus, $$\sum_{i=1}^nE\left|\mathbf{Z}_i\right|^{2+\delta}\leq O_p\left(\frac{p^{\delta/2}}{r^{\delta/2}}\right)\to0.$$ 


\noindent\begin{lemm}\label{lemm8}
 Under assumptions (C1)-(C7), if ${p_n^2}/{r}=o(1)$, we have $\bar{\mathbf{H}}_r\left(\boldsymbol{\beta}_{0}\right)-\bar{\mathbf{H}}_n\left(\boldsymbol{\beta}_{0}\right)=o_p\left(1\right).$
\end{lemm}

\noindent\textbf{\emph{The proof of Lemma \ref{lemm8}.}}   
Let $\bar{\mathbf{H}}_n\left(\boldsymbol{\beta}_{0}\right)=\frac{1}{n}\sum_{i=1}^n\mathbf{X}_i^T\mathbf{A}_i^{1/2}\left(\boldsymbol{\beta}_{0}\right)\bar{\mathbf{R}}^{-1}\mathbf{A}_i^{1/2}\left(\boldsymbol{\beta}_{0}\right)\mathbf{X}_i.$ Then  
\[
\bar{\mathbf{H}}_r(\boldsymbol{\beta}_{0}) - \bar{\mathbf{H}}_n(\boldsymbol{\beta}_{0}) = \frac{1}{n}\sum_{i=1}^n (\frac{\delta_i}{\pi_i}-1)\mathbf{X}_i^T \mathbf{A}_i^{1/2}(\boldsymbol{\beta}_{0})\bar{\mathbf{R}}^{-1}\mathbf{A}_i^{1/2}(\boldsymbol{\beta}_{0})\mathbf{X}_i.
\]
Note that 
$$\begin{aligned}
E\left(\bar{\mathbf{H}}_r\left(\boldsymbol{\beta}_{0}\right)\right) & =E\left[\frac{1}{n}\sum_{i=1}^n\frac{\delta_i}{\pi_i}\mathbf{X}_i^T \mathbf{A}_i^{1/2}(\boldsymbol{\beta}_{0})\bar{\mathbf{R}}^{-1}\mathbf{A}_i^{1/2}(\boldsymbol{\beta}_{0})\mathbf{X}_i\right] \\
 & =\frac{1}{n}\sum_{i=1}^n\mathbf{X}_i^T \mathbf{A}_i^{1/2}(\boldsymbol{\beta}_{0})\bar{\mathbf{R}}^{-1}\mathbf{A}_i^{1/2}(\boldsymbol{\beta}_{0})\mathbf{X}_i \\
 & =\bar{\mathbf{H}}_n(\boldsymbol{\beta}_{0}).
\end{aligned}$$
and
$$\begin{aligned}
Var(\bar{\mathbf{H}}_r\left(\boldsymbol{\beta}_{0}\right)) & =\frac{1}{n^2}\sum_{i=1}^n\frac{1-\pi_i}{\pi_i}\mathbf{X}_i^T \mathbf{A}_i^{1/2}(\boldsymbol{\beta}_{0})\bar{\mathbf{R}}^{-1}\mathbf{A}_i^{1/2}(\boldsymbol{\beta}_{0})\mathbf{X}_i\mathbf{X}_i^T\mathbf{A}_i^{1/2}\left(\boldsymbol{\beta}_{0}\right)\bar{\mathbf{R}}^{-1}\mathbf{A}_i^{1/2}\left(\boldsymbol{\beta}_{0}\right)\mathbf{X}_i \\
 & \leq\frac{1}{n^2}\sum_{i=1}^n\frac{1}{\pi_i}\mathbf{X}_i^T \mathbf{A}_i^{1/2}(\boldsymbol{\beta}_{0})\bar{\mathbf{R}}^{-1}\mathbf{A}_i^{1/2}(\boldsymbol{\beta}_{0})\mathbf{X}_i\mathbf{X}_i^T\mathbf{A}_i^{1/2}\left(\boldsymbol{\beta}_{0}\right)\bar{\mathbf{R}}^{-1}\mathbf{A}_i^{1/2}\left(\boldsymbol{\beta}_{0}\right)\mathbf{X}_i \\
 & \leq\left(\max_{1\leq i\leq n}\frac{1}{n\pi_i}\right)\frac{1}{n}\sum_{i=1}^n\left\|\mathbf{X}_i^T\mathbf{X}_i\right\|^2 \\
 & =o_p
\begin{pmatrix}
1
\end{pmatrix}.
\end{aligned}$$
Applying Chebyshev's inequality, the lemma \ref{lemm8} is thus proved.

\bmsection{The proof of Theorem}
\noindent\textbf{\emph{The proof of Theorem \ref{theorem1} } } We now show that, for any \(\epsilon > 0\), there exists a constant \(\Delta > 0\) such that when \(r\) is sufficiently large, $$P\left(\sup_{\left\|\boldsymbol{\beta}-\boldsymbol{\beta}_{0}\right\|=\Delta\sqrt{{p_n}/{r}}}\left(\boldsymbol{\beta}-\boldsymbol{\beta}_{0}\right)^T\mathbf{S}_r\left(\boldsymbol{\beta}\right)<0\right)\geq1-\epsilon.$$
Note 
$$
\begin{aligned}
(\boldsymbol{\beta}-\boldsymbol{\beta}_{0})^T S_r(\boldsymbol{\beta}) &= (\boldsymbol{\beta}-\boldsymbol{\beta}_{0})^T \mathbf{S}_r(\boldsymbol{\beta}_{0}) - (\boldsymbol{\beta}-\boldsymbol{\beta}_{0})^T D_r(\boldsymbol{\beta}^*) (\boldsymbol{\beta}-\boldsymbol{\beta}_{0}) \\
&\triangleq I_{n1} + I_{n2},
\end{aligned}
$$
where $\mathbf{D}_r\left(\boldsymbol{\beta}^*\right)=-\frac{\partial \mathbf{S}_r\left(\boldsymbol{\beta}^*\right)}{\partial\boldsymbol{\beta}},\|\boldsymbol{\beta}^*-\boldsymbol{\beta}\|\leq\|\boldsymbol{\beta}_{0}-\boldsymbol{\beta}\|$. Next, we have 
$$\begin{aligned}
I_{n1} & =\left(\boldsymbol{\beta}-\boldsymbol{\beta}_{0}\right)^T\bar{\mathbf{S}}_r\left(\boldsymbol{\beta}_{0}\right)+\left(\boldsymbol{\beta}-\boldsymbol{\beta}_{0}\right)^T\left[\mathbf{S}_r\left(\boldsymbol{\beta}_{0}\right)-\bar{\mathbf{S}}_r\left(\boldsymbol{\beta}_{0}\right)\right] \\
 & \triangleq I_{n11}+I_{n12}.
\end{aligned}$$
By the Cauchy-Schwarz inequality, we have $\left|I_{n11}\right|\leq\|\boldsymbol{\beta}-\boldsymbol{\beta}_{0}\|\|\bar{\mathbf{S}}_r\left(\boldsymbol{\beta}_{0}\right)\|=\Delta\sqrt{{p_n}/{r}}\left\|\bar{\mathbf{S}}_r\left(\boldsymbol{\beta}_{0}\right)\right\|$. Furthermore, 
$$\begin{aligned}
E\left[\left\|\bar{\mathbf{S}}_r\left(\boldsymbol{\beta}_{0}\right)\right\|^2\right] & =E\left\{\frac{1}{n^2}\sum_{i=1}^n\frac{1}{\pi_i}\boldsymbol{\varepsilon}_i^T\left(\boldsymbol{\beta}_{0}\right)\bar{\mathbf{R}}^{-1}\mathbf{A}_i^{1/2}\left(\boldsymbol{\beta}_{0}\right)\mathbf{X}_i\mathbf{X}_i^T\mathbf{A}_i^{1/2}\left(\boldsymbol{\beta}_{0}\right)\bar{\mathbf{R}}^{-1}\boldsymbol{\varepsilon}_i\left(\boldsymbol{\beta}_{0}\right)\right\} \\
 & \leq\frac{1}{n^2}\sum_{i=1}^n\frac{1}{\pi_i}\lambda_{\max}(\bar{\mathbf{R}}^{-2})\lambda_{\max}\left(\mathbf{X}_i\mathbf{X}_i^T\right)E\left[\boldsymbol{\varepsilon}_i^T\left(\boldsymbol{\beta}_{0}\right)\boldsymbol{\varepsilon}_i\left(\boldsymbol{\beta}_{0}\right)\right] \\
 & \leq C\left(\max_{1\leq i\leq n}\frac{1}{n\pi_i}\right)Tr\left(\frac{1}{n}\sum_{i=1}^n\mathbf{X}_i\mathbf{X}_i^T\right) \\
 & =O_p\left({p_n}/{r}\right).
\end{aligned}$$ 
Therefore, $\left\|\bar{\mathbf{S}}_r\left(\boldsymbol{\beta}_{0}\right)\right\|=O_p(\sqrt{{p_n}/{r}})$. Consequently, $\left|I_{n1}\right|=\Delta O_p\left({p_n}/{r}\right)$. By lemma \ref{lemm2}, we can derive that $I_{n12}\leq\left\|\boldsymbol{\beta}-\boldsymbol{\beta}_{0}\right\|\|\mathbf{S}_r\left(\boldsymbol{\beta}_{0}\right)-\bar{\mathbf{S}}_r\left(\boldsymbol{\beta}_{0}\right)\|=\Delta\sqrt{{p_n}/{r}}O_p\left({p_n}/{r}\right)=\Delta o_p\left({p_n}/{r}\right)$.  Next, we disuass $I_{n2}$, 
$$\begin{aligned}
I_{n2} & =-\left(\boldsymbol{\beta}-\boldsymbol{\beta}_{0}\right)^T\bar{\mathbf{D}}_r\left(\boldsymbol{\beta}^*\right)\left(\boldsymbol{\beta}-\boldsymbol{\beta}_{0}\right)-\left(\boldsymbol{\beta}-\boldsymbol{\beta}_{0}\right)^T\left[\mathbf{D}_r\left(\boldsymbol{\beta}^*\right)-\bar{\mathbf{D}}_r\left(\boldsymbol{\beta}^*\right)\right]\left(\boldsymbol{\beta}-\boldsymbol{\beta}_{0}\right) \\
 & \triangleq I_{n21}+I_{n22}.
\end{aligned}$$
Note that by Lemma \ref{lemm5}, we have 
$$\begin{aligned}
\left|I_{n22}\right| & \leq\max\left(\left|\lambda_{\max}\left(\mathbf{D}_r\left(\boldsymbol{\beta}^*\right)-\bar{\mathbf{D}}_r\left(\boldsymbol{\beta}^*\right)\right)\right|,\left|\lambda_{\min}\left(\mathbf{D}_r\left(\boldsymbol{\beta}^*\right)-\bar{\mathbf{D}}_r\left(\boldsymbol{\beta}^*\right)\right)\right|\right)\left\|\boldsymbol{\beta}-\boldsymbol{\beta}_{0}\right\|^2 \\
 & =O_p(\sqrt{{p_n}/{r}})\Delta^2{p_n}/{r} \\
 & =\Delta^2o_p({p_n}/{r}).
\end{aligned}$$
For $I_{n2},$ we have 
$$
\begin{aligned}
I_{n21} &= -\left(\boldsymbol{\beta} - \boldsymbol{\beta}_{0}\right)^T \left[\bar{\mathbf{H}}_r\left(\boldsymbol{\beta}^*\right) + \bar{\mathbf{E}}_r\left(\boldsymbol{\beta}^*\right) + \bar{\mathbf{G}}_r\left(\boldsymbol{\beta}^*\right)\right] \left(\boldsymbol{\beta} - \boldsymbol{\beta}_{0}\right) \\
&= -\left(\boldsymbol{\beta} - \boldsymbol{\beta}_{0}\right)^T \bar{\mathbf{H}}_r\left(\boldsymbol{\beta}_{0}\right) \left(\boldsymbol{\beta} - \boldsymbol{\beta}_{0}\right) \\
&\quad -\left(\boldsymbol{\beta} - \boldsymbol{\beta}_{0}\right)^T \left[\bar{\mathbf{H}}_r\left(\boldsymbol{\beta}^*\right) - \bar{\mathbf{H}}_r\left(\boldsymbol{\beta}_{0}\right)\right] \left(\boldsymbol{\beta} - \boldsymbol{\beta}_{0}\right) \\
&\quad -\left(\boldsymbol{\beta} - \boldsymbol{\beta}_{0}\right)^T \bar{\mathbf{E}}_r\left(\boldsymbol{\beta}^*\right) \left(\boldsymbol{\beta} - \boldsymbol{\beta}_{0}\right) \\
&\quad -\left(\boldsymbol{\beta} - \boldsymbol{\beta}_{0}\right)^T \bar{\mathbf{G}}_r\left(\boldsymbol{\beta}^*\right) \left(\boldsymbol{\beta} - \boldsymbol{\beta}_{0}\right) \\
&= I_{n21}^a + I_{n21}^b + I_{n21}^c + I_{n21}^d.
\end{aligned}
$$
By Lemmas \ref{lemm5} and \ref{lemm6}, $I_{n21}^b=\Delta^2{p}/{r}O_p\left(\sqrt{{p_n}/{r}}\right)=\Delta^2o_p\left({p_n}/{r}\right),I_{n21}^c=I_{n21}^d=\Delta^2{p_n}/{r}
O_p\left({p_n}/{\sqrt{r}}\right)=\Delta^2o_p\left({p_n}/{r}\right).$ 
Finally, under assumptions (C3) and (C7), we analyze $I_{n21}^a,$ 
$$\begin{aligned}
I_{n21}^{a} & =-\left(\boldsymbol{\beta}-\boldsymbol{\beta}_{0}\right)^T\left[\frac{1}{n}\sum_{i=1}^n\frac{\delta_i}{\pi_i}\mathbf{X}_i^T\mathbf{A}_i^{1/2}\left(\boldsymbol{\beta}_{0}\right)\bar{\mathbf{R}}^{-1}\mathbf{A}_i^{1/2}\left(\boldsymbol{\beta}_{0}\right)\mathbf{X}_i\right]\left(\boldsymbol{\beta}-\boldsymbol{\beta}_{0}\right) \\
 & \leq-\lambda_{\min}\left(\bar{\mathbf{R}}^{-1}\right)\min_i\lambda_{\min}\left(\mathbf{A}_i\left(\boldsymbol{\beta}_{0}\right)\right)\lambda_{\min}\left(\frac{1}{n}\sum_{i=1}^n\frac{\delta_i}{\pi_i}\mathbf{X}_i^T\mathbf{X}_i\right)\left\|\boldsymbol{\beta}-\boldsymbol{\beta}_{0}\right\|^2 \\
 & \leq-C\Delta^2{p_n}/{r}.
\end{aligned}$$
Therefore, for $\left(\boldsymbol{\beta}-\boldsymbol{\beta}_{0}\right)^TS_r\left(\boldsymbol{\beta}\right)$ on the set $\mathcal{B}=\left\{\boldsymbol{\beta}{:}\left\|\boldsymbol{\beta}-\boldsymbol{\beta}_{0}\right\|\leq\Delta\sqrt{{p_n}/{r}}\right\}$, it is dominated by $I_{n11}$ and $I_{n21}^a$. When $\Delta >0$ is sufficiently large, it can be negative. This finishes the proof of Theorem \ref{theorem1}.  \\


\noindent\textbf{\emph{The proof of Theorem \ref{theorem2} }} 
First, we prove 
\begin{equation}\label{A6}
\mathbf{c}_n^T \bar{\mathbf{M}}_r^{-1/2}(\boldsymbol{\beta}_{0}) \bar{\mathbf{H}}_r(\boldsymbol{\beta}_{0}) (\tilde{\boldsymbol{\beta}} - \boldsymbol{\beta}_{0}) \overset{d}{\longrightarrow} N(0,1).
\end{equation}
We have 
$$\begin{aligned}
\mathbf{c}_n^T\bar{\mathbf{M}}_r^{-1/2}\left(\boldsymbol{\beta}_{0}\right)\bar{\mathbf{S}}_r\left(\boldsymbol{\beta}_{0}\right) & =\mathbf{c}_n^T\bar{\mathbf{M}}_r^{-1/2}\left(\boldsymbol{\beta}_{0}\right)\mathbf{S}_r\left(\boldsymbol{\beta}_{0}\right)+\mathbf{c}_n^T\bar{\mathbf{M}}_r^{-1/2}\left(\boldsymbol{\beta}_{0}\right)\left[\bar{\mathbf{S}}_r\left(\boldsymbol{\beta}_{0}\right)-\mathbf{S}_r\left(\boldsymbol{\beta}_{0}\right)\right] \\
 & =\mathbf{c}_n^T\bar{\mathbf{M}}_r^{-1/2}\left(\boldsymbol{\beta}_{0}\right)\mathbf{D}_r\left(\boldsymbol{\beta}^*\right)(\tilde{\boldsymbol{\beta}}-\boldsymbol{\beta}_{0})+\mathbf{c}_n^T\bar{\mathbf{M}}_r^{-1/2}\left(\boldsymbol{\beta}_{0}\right)\left[\bar{\mathbf{S}}_r\left(\boldsymbol{\beta}_{0}\right)-\mathbf{S}_r\left(\boldsymbol{\beta}_{0}\right)\right] \\
 & =\mathbf{c}_n^T\bar{\mathbf{M}}_r^{-1/2}\left(\boldsymbol{\beta}_{0}\right)\bar{\mathbf{H}}_r\left(\boldsymbol{\beta}_{0}\right)(\tilde{\boldsymbol{\beta}}-\boldsymbol{\beta}_{0})\\
 & \quad
+\mathbf{c}_n^T\bar{\mathbf{M}}_r^{-1/2}\left(\boldsymbol{\beta}_{0}\right)\left[\mathbf{D}_r\left(\boldsymbol{\beta}^*\right)-\bar{\mathbf{H}}_r\left(\boldsymbol{\beta}_{0}\right)\right](\tilde{\boldsymbol{\beta}}-\boldsymbol{\beta}_{0}) \\
 & \quad
+\mathbf{c}_n^T\bar{\mathbf{M}}_r^{-1/2}\left(\boldsymbol{\beta}_{0}\right)[\bar{\mathbf{S}}_r\left(\boldsymbol{\beta}_{0}\right)-\mathbf{S}_r\left(\boldsymbol{\beta}_{0}\right)].
\end{aligned}$$
In the second equality, since $\mathbf{S}_r(\tilde{\boldsymbol{\beta}})=0$, Taylor expansion yields $$\mathbf{S}_r\left(\boldsymbol{\beta}_{0}\right)=\mathbf{D}_r\left(\boldsymbol{\beta}^*\right)(\tilde{\boldsymbol{\beta}}-\boldsymbol{\beta}_{0}),\left\|\boldsymbol{\beta}^*-\boldsymbol{\beta}_{0}\right\|\leq\|\tilde{\boldsymbol{\beta}}-\boldsymbol{\beta}_{0}\|.$$ By Lemma \ref{lemm7}, 
$\mathbf{c}_n^T \bar{\mathbf{M}}_r^{-1/2}(\boldsymbol{\beta}_{0}) \bar{\mathbf{S}}_r(\boldsymbol{\beta}_{0}) (\tilde{\boldsymbol{\beta}} - \boldsymbol{\beta}_{0}) \overset{d}{\longrightarrow} N(0,1).$ Thus, to prove (\ref{A6}), it suffices to show that for any $\Delta >0$, 
\begin{equation}\label{A7}
\sup_{\|\boldsymbol{\beta} - \boldsymbol{\beta}_{0}\| \leq \Delta\sqrt{{p_n}/{r}}} \left|\mathbf{c}_n^T \bar{\mathbf{M}}_r^{-1/2}(\boldsymbol{\beta}_{0}) \left[\mathbf{D}_r(\boldsymbol{\beta}) - \bar{\mathbf{H}}_r(\boldsymbol{\beta})\right] (\tilde{\boldsymbol{\beta}} - \boldsymbol{\beta}_{0})\right| = o_p(1),
\end{equation}
and 
\begin{equation}\label{A8}
 \left|\mathbf{c}_n^T \bar{\mathbf{M}}_r^{-1/2}(\boldsymbol{\beta}_{0}) \left[\bar{\mathbf{S}}_r(\boldsymbol{\beta}_{0}) - {\mathbf{S}}_r(\boldsymbol{\beta}_{0})\right] \right| = o_p(1).
\end{equation}
First, we prove (\ref{A8}). By Lemma \ref{lemm2} and (\ref{A5}), observe that 
\begin{align*}
&\left[\mathbf{c}_{n}^{T} \bar{\mathbf{M}}_r^{-1/2}(\boldsymbol{\beta}_{0})\left[\bar{\mathbf{S}}_r(\boldsymbol{\beta}_{0})-\mathbf{S}_r(\boldsymbol{\beta}_{0})\right]\right]^{2} \\
&= \mathbf{c}_{n}^{T} \bar{\mathbf{M}}_r^{-1/2}(\boldsymbol{\beta}_{0})\left[\bar{\mathbf{S}}_r(\boldsymbol{\beta}_{0})-\mathbf{S}_r(\boldsymbol{\beta}_{0})\right]\left[\bar{\mathbf{S}}_r(\boldsymbol{\beta}_{0})-\mathbf{S}_r(\boldsymbol{\beta}_{0})\right]^{T} \bar{\mathbf{M}}_r^{-1/2}(\boldsymbol{\beta}_{0}) \mathbf{c}_{n} \\
&\leq \lambda_{\max}\left(\bar{\mathbf{M}}_r^{-1}(\boldsymbol{\beta}_{0})\right) \lambda_{\max}\left(\left[\bar{\mathbf{S}}_r(\boldsymbol{\beta}_{0})-\mathbf{S}_r(\boldsymbol{\beta}_{0})\right]\left[\bar{\mathbf{S}}_r(\boldsymbol{\beta}_{0})-\mathbf{S}_r(\boldsymbol{\beta}_{0})\right]^{T}\right) \\
&\leq \frac{\left\|\bar{\mathbf{S}}_r(\boldsymbol{\beta}_{0})-\mathbf{S}_r(\boldsymbol{\beta}_{0})\right\|^{2}}{\lambda_{\min}\left(\bar{\mathbf{M}}_r(\boldsymbol{\beta}_{0})\right)} \\
&\leq \frac{\left\|\bar{\mathbf{S}}_r(\boldsymbol{\beta}_{0})-\mathbf{S}_r(\boldsymbol{\beta}_{0})\right\|^{2}}{C \lambda_{\min}\left(\frac{1}{n} \sum_{i=1}^{n} \frac{1}{n \pi_{i}} X_{i}^{T} X_{i}\right)} \\
&= O_{p}\left({p_{n}^{2}}/{r}\right) = o_{p}(1).
\end{align*}
Next, we prove (\ref{A7}),
\begin{align*}
&\sup_{\|\boldsymbol{\beta}-\boldsymbol{\beta}_{0}\|\leq\Delta\sqrt{{p_n}/{r}}}\left|\mathbf{c}_n^T\bar{\mathbf{M}}_r^{-1/2}(\boldsymbol{\beta}_{0})\left[\mathbf{D}_r(\boldsymbol{\beta})-\bar{\mathbf{H}}_r(\boldsymbol{\beta}_{0})\right](\tilde{\boldsymbol{\beta}}-\boldsymbol{\beta}_{0})\right| \\
& \leq \sup_{\|\boldsymbol{\beta}-\boldsymbol{\beta}_{0}\|\leq\Delta\sqrt{{p_n}/{r}}}\left|\mathbf{c}_n^T\bar{\mathbf{M}}_r^{-1/2}(\boldsymbol{\beta}_{0})\left[\mathbf{D}_r(\boldsymbol{\beta})-\bar{\mathbf{D}}_r(\boldsymbol{\beta})\right](\tilde{\boldsymbol{\beta}}-\boldsymbol{\beta}_{0})\right| \\
&\quad + \sup_{\|\boldsymbol{\beta}-\boldsymbol{\beta}_{0}\|\leq\Delta\sqrt{{p_n}/{r}}}\left|\mathbf{c}_n^T\bar{\mathbf{M}}_r^{-1/2}(\boldsymbol{\beta}_{0})\left[\bar{\mathbf{D}}_r(\boldsymbol{\beta})-\bar{\mathbf{H}}_r(\boldsymbol{\beta})\right](\tilde{\boldsymbol{\beta}}-\boldsymbol{\beta}_{0})\right| \\
&\quad + \sup_{\|\boldsymbol{\beta}-\boldsymbol{\beta}_{0}\|\leq\Delta\sqrt{{p_n}/{r}}}\left|\mathbf{c}_n^T\bar{\mathbf{M}}_r^{-1/2}(\boldsymbol{\beta}_{0})\left[\bar{\mathbf{H}}_r(\boldsymbol{\beta})-\bar{\mathbf{H}}_r(\boldsymbol{\beta}_{0})\right](\tilde{\boldsymbol{\beta}}-\boldsymbol{\beta}_{0})\right| \\
& \triangleq I_{n1} + I_{n2} + I_{n3}.
\end{align*}
By the Cauchy-Schwarz inequality and Lemma \ref{lemm5}, we have
$$\begin{aligned}
I_{n1} & \leq\sup_{\left\|\boldsymbol{\beta}-\boldsymbol{\beta}_{0}\right\|\leq\Delta\sqrt{{p_n}/{r}}}\left[\mathbf{c}_n^T\bar{\mathbf{M}}_r^{-1/2}\left(\boldsymbol{\beta}_{0}\right)\left(\mathbf{D}_r\left(\boldsymbol{\beta}\right)-\bar{\mathbf{D}}_r\left(\boldsymbol{\beta}\right)\right)^2\bar{\mathbf{M}}_r^{-1/2}\left(\boldsymbol{\beta}_{0}\right)\mathbf{c}_n\right]^{1/2}\left\|\tilde{\boldsymbol{\beta}}-\boldsymbol{\beta}_{0}\right\| \\
 & \leq\sup_{\|\boldsymbol{\beta}-\boldsymbol{\beta}_{0}\|\leq\Delta\sqrt{{p_n}/{r}}}\max\left(\left|\lambda_{\min}\left(\mathbf{D}_r\left(\boldsymbol{\beta}\right)-\bar{\mathbf{D}}_r\left(\boldsymbol{\beta}\right)\right)\right|,\left|\lambda_{\max}\left(\mathbf{D}_r\left(\boldsymbol{\beta}\right)-\bar{\mathbf{D}}_r\left(\boldsymbol{\beta}\right)\right)\right|\right)\\
&\quad \times\lambda_{\min}^{-1/2}(\bar{\mathbf{M}}_r\left(\boldsymbol{\beta}_{0}\right))O_p(\sqrt{{p_n}/{r}}) \\
 & =O_p(\sqrt{{p_n^2}/{r}})=o_p\left(1\right).
\end{aligned}$$
By Lemmas \ref{lemm5} and \ref{lemm6}, we can also derive $I_{n2}=I_{n3}=o_{p} (1)$. Thus, (\ref{A6}) is proved.
By Lemma \ref{lemm8}, we have $\mathbf{c}_n^T\bar{\mathbf{M}}_r^{-1/2}\left(\boldsymbol{\beta}_{0}\right)\left[\bar{\mathbf{H}}_r\left(\boldsymbol{\beta}_{0}\right)-\bar{\mathbf{H}}_n\left(\boldsymbol{\beta}_{0}\right)\right](\tilde{\boldsymbol{\beta}}-\boldsymbol{\beta}_{0})=o_p\left(1\right)$. Combining this with Slutsky's theorem yields $\mathbf{c}_n^T \bar{\mathbf{M}}_r^{-1/2}(\boldsymbol{\beta}_{0}) \bar{\mathbf{H}}_n(\boldsymbol{\beta}_{0}) (\tilde{\boldsymbol{\beta}} - \boldsymbol{\beta}_{0}) \overset{d}{\longrightarrow} N(0,1).$ This finishes the proof of Theorem \ref{theorem2}.\\


\noindent\textbf{\emph{The proof of Theorem \ref{theorem3} }} 
It suffices to show that for any $\mathbf{c}_n\in\mathbf{R}^{p_n}$, \begin{equation}\label{A9}
\sup_{\|\mathbf{c}_n\|=1} \left|\mathbf{c}_n^T (\tilde{\boldsymbol{\Sigma}} - \boldsymbol{\Sigma}) \mathbf{c}_n\right| = o_p(1).
\end{equation}
In our proof, we employ Theorem \ref{theorem1}. Let $\tilde{\boldsymbol{\Sigma}} - \boldsymbol{\Sigma}=I_{n1}+I_{n2}+I_{n3}$, where 
$$\begin{aligned}
 & \tilde{\boldsymbol{\Sigma}}=\mathbf{H}_n^{-1}(\tilde{\boldsymbol{\beta}})\tilde{\mathbf{M}}_r(\tilde{\boldsymbol{\beta}})\mathbf{H}_n^{-1}(\tilde{\boldsymbol{\beta}}). \\
 & I_{n1}=\mathbf{H}_n^{-1}(\tilde{\boldsymbol{\beta}})\left[\tilde{\mathbf{M}}_r(\tilde{\boldsymbol{\beta}})-\bar{\mathbf{M}}_r(\boldsymbol{\beta}_{0})\right]\mathbf{H}_n^{-1}(\tilde{\boldsymbol{\beta}}), \\
 & I_{n2}=\left[\mathbf{H}_n^{-1}(\tilde{\boldsymbol{\beta}})-\bar{\mathbf{H}}_n^{-1}\left(\boldsymbol{\beta}_{0}\right)\right]\bar{\mathbf{M}}_r\left(\boldsymbol{\beta}_{0}\right)\mathbf{H}_n^{-1}(\tilde{\boldsymbol{\beta}}), \\
 & I_{n3}=\bar{\mathbf{H}}_{n}^{-1}\left(\boldsymbol{\beta}_{0}\right)\bar{\mathbf{M}}_r\left(\boldsymbol{\beta}_{0}\right)\left[\mathbf{H}_{n}^{-1}(\tilde{\boldsymbol{\beta}})-\bar{\mathbf{H}}_{n}^{-1}\left(\boldsymbol{\beta}_{0}\right)\right].
\end{aligned}$$
Therefore, (\ref{A9}) can be derived from $\sup_{\|\mathbf{c}_n\|=1}\left|\mathbf{c}_n^T\mathbf{I}_{ni}\mathbf{c}_n\right|=o_p\left(1\right)$. Furthermore, we have 
$$\sup_{\left\|\mathbf{c}_n\right\|=1}\left|\mathbf{c}_n^T\mathbf{I}_{n1}\mathbf{c}_n\right|\leq\frac{\max\left(\left|\lambda_{\min}(\tilde{\mathbf{M}}_r(\tilde{\boldsymbol{\beta}})-\bar{\mathbf{M}}_r\left(\boldsymbol{\beta}_{0}\right))\right|,\left|\lambda_{\max}(\tilde{\mathbf{M}}_r(\tilde{\boldsymbol{\beta}})-\bar{\mathbf{M}}_r\left(\boldsymbol{\beta}_{0}\right))\right|\right)}{\lambda_{\min}^2(\mathbf{H}_n(\tilde{\boldsymbol{\beta}}))}.$$
To analyze the eigenvalues of $\tilde{\mathbf{M}}_r(\tilde{\boldsymbol{\beta}})-\bar{\mathbf{M}}_r\left(\boldsymbol{\beta}_{0}\right)$, 
$$\begin{aligned}
 & \left|\mathbf{c}_n^T\left[\tilde{\mathbf{M}}_r(\tilde{\boldsymbol{\beta}})-\bar{\mathbf{M}}_r\left(\boldsymbol{\beta}_{0}\right)\right]\mathbf{c}_n\right| \\
 & \leq\left|\mathbf{c}_n^T\left[\tilde{\mathbf{M}}_r(\tilde{\boldsymbol{\beta}})-\tilde{\mathbf{M}}_r\left(\boldsymbol{\beta}_{0}\right)\right]\mathbf{c}_n\right|+\left|\mathbf{c}_n^T\left[\tilde{\mathbf{M}}_r\left(\boldsymbol{\beta}_{0}\right)-\bar{\mathbf{M}}_r\left(\boldsymbol{\beta}_{0}\right)\right]\mathbf{c}_n\right|.
\end{aligned}$$
Note that 
$$\begin{aligned}
 & \sup_{\|\mathbf{c}_n=1\|}\left|\mathbf{c}_n^T\left[\tilde{\mathbf{M}}_r(\tilde{\boldsymbol{\beta}})-\tilde{\mathbf{M}}_r\left(\boldsymbol{\beta}_{0}\right)\right]\mathbf{c}_n\right| \\
 & \leq\sup_{\|\mathbf{c}_n=1\|}\left|\frac{1}{n}\sum_{i=1}^n\frac{\delta_i}{\pi_i}\mathbf{c}_n^T\mathbf{X}_i^T\left[\mathbf{A}_i^{1/2}(\tilde{\boldsymbol{\beta}})-\mathbf{A}_i^{1/2}(\boldsymbol{\beta}_{0})\right]\tilde{\mathbf{R}}^{-1}\boldsymbol{\varepsilon}_i(\tilde{\boldsymbol{\beta}})\boldsymbol{\varepsilon}_i^T(\tilde{\boldsymbol{\beta}})\tilde{\mathbf{R}}^{-1}\mathbf{A}_i^{1/2}(\tilde{\boldsymbol{\beta}})\mathbf{X}_i\mathbf{c}_n\right| \\
 & \quad +\sup_{\|\mathbf{c}_n=1\|}\left|\frac{1}{n}\sum_{i=1}^n\frac{\delta_i}{\pi_i}\mathbf{c}_n^T\mathbf{X}_i^T\mathbf{A}_i^{1/2}\left(\boldsymbol{\beta}_{0}\right)\tilde{\mathbf{R}}^{-1}\boldsymbol{\varepsilon}_i(\tilde{\boldsymbol{\beta}})\boldsymbol{\varepsilon}_i^T(\tilde{\boldsymbol{\beta}})\tilde{\mathbf{R}}^{-1}\left[\mathbf{A}_i^{1/2}(\tilde{\boldsymbol{\beta}})-\mathbf{A}_i^{1/2}\left(\boldsymbol{\beta}_{0}\right)\right]\mathbf{X}_i\mathbf{c}_n\right| \\
 & \quad  +\sup_{\left\|\mathbf{c}_n=1\right\|}\left|\frac{1}{n}\sum_{i=1}^n\frac{\delta_i}{\pi_i}\mathbf{c}_n^T\mathbf{X}_i^T\mathbf{A}_i^{1/2}\left(\boldsymbol{\beta}_{0}\right)\tilde{\mathbf{R}}^{-1}\left[\boldsymbol{\varepsilon}_i(\tilde{\boldsymbol{\beta}})\boldsymbol{\varepsilon}_i^T(\tilde{\boldsymbol{\beta}})-\boldsymbol{\varepsilon}_i(\boldsymbol{\beta}_{0})\boldsymbol{\varepsilon}_i^T(\boldsymbol{\beta}_{0})\right]\tilde{\mathbf{R}}^{-1}\mathbf{A}_i^{1/2}\left(\boldsymbol{\beta}_{0}\right)\mathbf{X}_i\mathbf{c}_n\right| \\
 & \triangleq\sup_{\left\|\mathbf{c}_n=1\right\|}J_{n1}+\sup_{\left\|\mathbf{c}_n=1\right\|}J_{n2}+\sup_{\left\|\mathbf{c}_n=1\right\|}J_{n3}.
\end{aligned}$$
It can see
$$
J_{n1} \leq \frac{1}{n}\sum_{i=1}^n \frac{\delta_i}{\pi_i}
\left\|\mathbf{c}_n^T\mathbf{X}_i^T \left[\mathbf{A}_i^{1/2}(\tilde{\boldsymbol{\beta}}) - \mathbf{A}_i^{1/2}(\boldsymbol{\beta}_{0})\right]\right\|
\left\|\tilde{\mathbf{R}}^{-1}\boldsymbol{\varepsilon}_i(\tilde{\boldsymbol{\beta}})\right\|^2
\left\|\mathbf{A}_i^{1/2}(\tilde{\boldsymbol{\beta}})\mathbf{X}_i\mathbf{c}_n\right\|.
$$
We have
$\|\mathbf{A}_i^{1/2}(\tilde{\boldsymbol{\beta}})\mathbf{X}_i\mathbf{c}_n\|\leq\left\|\mathbf{X}_i\mathbf{c}_n\right\|$ and
$$\begin{aligned}
\left\|\mathbf{c}_n^T\mathbf{X}_i^T\left[\mathbf{A}_i^{1/2}(\tilde{\boldsymbol{\beta}})-\mathbf{A}_i^{1/2}\left(\boldsymbol{\beta}_{0}\right)\right]\right\| & \leq\left\|\mathbf{X}_i\mathbf{c}_n\right\|\max_j\left|\mathbf{A}_{ij}^{1/2}(\tilde{\boldsymbol{\beta}})-\mathbf{A}_{ij}^{1/2}\left(\boldsymbol{\beta}_{0}\right)\right| \\
 & \leq C\left\|\mathbf{X}_i\mathbf{c}_n\right\|\cdot\left\|\mathbf{X}_{ij}\right\|\cdot\|\tilde{\boldsymbol{\beta}}-\boldsymbol{\beta}_{0}\|,
\end{aligned}$$
and 
$$\begin{aligned}
\left\|\tilde{\mathbf{R}}^{-1}\boldsymbol{\varepsilon}_i(\tilde{\boldsymbol{\beta}})\right\|^2 & =(\mathbf{Y}_i-\boldsymbol{\mu}_i(\tilde{\boldsymbol{\beta}}))^T\mathbf{A}_i^{-1/2}(\tilde{\boldsymbol{\beta}})\tilde{\mathbf{R}}^{-2}\mathbf{A}_i^{-1/2}(\tilde{\boldsymbol{\beta}})(\mathbf{Y}_i-\boldsymbol{\mu}_i(\tilde{\boldsymbol{\beta}})) \\
 & \leq\lambda_{\max}(\tilde{\mathbf{R}}^{-2})\lambda_{\max}(\mathbf{A}_i^{-1}(\tilde{\boldsymbol{\beta}}))\|\mathbf{Y}_i-\boldsymbol{\mu}_i(\tilde{\boldsymbol{\beta}})\|^2 \\
 & \leq CO_p(1).
\end{aligned}$$
Therefore, we have 
$$\sup_{\|c_n=1\|}J_{n1}\leq O_p\left(1\right)\|\tilde{\boldsymbol{\beta}}-\boldsymbol{\beta}_{0}\|\max_{i,j}\left\|\mathbf{X}_{ij}\right\|\lambda_{\max}\left(\frac{1}{n}\sum_{i=1}^n\frac{\delta_i}{\pi_i}\mathbf{X}_i^T\mathbf{X}_i\right)=o_p\left(1\right).$$
Similarly, $\sup_{\|\mathbf{c}_n=1\|}J_{n2}=o_p (1)$ and $\sup_{\|\mathbf{c}_n=1\|}J_{n3}=o_p (1)$. Hence, 
$$\sup_{\|\mathbf{c}_n\|=1}\lvert \mathbf{c}_n^T[\tilde{\mathbf{M}}_r(\tilde{\boldsymbol{\beta}})-\tilde{\mathbf{M}}_r(\boldsymbol{\beta}_{0})]\mathbf{c}_n\rvert=o_p\left(1\right).$$
Similarly, we have $\sup_{\|\mathbf{c}_n\|=1}|\mathbf{c}_n^T[\tilde{\mathbf{M}}_r\left(\boldsymbol{\beta}_{0}\right)-\bar{\mathbf{M}}_r\left(\boldsymbol{\beta}_{0}\right)]\mathbf{c}_n|=o_p\left(1\right)$. Finally, observe that 
$$\begin{aligned}
\lambda_{\min}(\mathbf{H}_{n}(\tilde{\boldsymbol{\beta}})) & \geq\lambda_{\min}(\tilde{\mathbf{R}}^{-1})\lambda_{\min}\left(\frac{1}{n}\sum_{i=1}^n\mathbf{X}_i^T\mathbf{X}_i\right) \\
 & =O_p(1).
\end{aligned}$$
Therefore, we have $\sup_{\|\mathbf{b}_n\|=1}\left|\mathbf{b}_n^T\mathbf{I}_{n1}\mathbf{b}_n\right|=o_p\left(1\right)$. Using 
$$\mathbf{H}_{n}^{-1}(\tilde{\boldsymbol{\beta}})-\bar{\mathbf{H}}_{n}^{-1}(\boldsymbol{\beta}_0)=[\mathbf{H}_{n}^{-1}(\tilde{\boldsymbol{\beta}})-\bar{\mathbf{H}}_{n}^{-1}(\tilde{\boldsymbol{\beta}})]+[\bar{\mathbf{H}}_{n}^{-1}(\tilde{\boldsymbol{\beta}})-\bar{\mathbf{H}}_{n}^{-1}(\boldsymbol{\beta}_0)],$$
we also obtain $\sup_{\|\mathbf{b}_n\|=1}\left|\mathbf{b}_n^T\mathbf{I}_{ni}\mathbf{b}_n\right|=o_p\left(1\right),i=2,3.$ This finishes the proof of Theorem \ref{theorem3}.

\noindent\textbf{\emph{The proof of Theorem \ref{tthree} and \ref{tfour} } } If some elements of \( \left\{h_i\right\}_{i=1}^n \) equal zero, their associated subsampling probabilities are set to zero, and the subsampling probabilities of the remaining individuals are considered.  
Thus, we assume all \( h_i > 0 \), which does not affect generality. 

To minimize \(tr(\bar{\mathbf{H}}_n^{-1}(\boldsymbol{\beta}_{0})\bar{\mathbf{M}}_r(\boldsymbol{\beta}_{0})\bar{\mathbf{H}}_n^{-1}(\boldsymbol{\beta}_{0}))\), which corresponds to the asymptotic mean squared error, the following optimization problem needs to be solved:  
\begin{equation}
\begin{aligned} \label{24}
& \min \quad \tilde{H} = \sum_{i=1}^n tr \left[\frac{1}{\pi_i} \left\|\bar{\mathbf{H}}_{n}^{-1}\left(\boldsymbol{\beta}_{0}\right)\mathbf{X}_i^T\mathbf{A}_i^{1/2}\left(\boldsymbol{\beta}_{0}\right)\bar{\mathbf{R}}^{-1}\boldsymbol{\varepsilon}_i\left(\boldsymbol{\beta}_{0}\right)\right\|^2 \right] \\ 
& \operatorname{s.t.} \quad \sum_{i=1}^n \pi_i = r, \quad 0 \leq \pi_i \leq 1, \quad i = 1, \dots, n.
\end{aligned}
\end{equation}

For simplicity, define \( h_i \) as  
$h_i^{MV} = \|\bar{\mathbf{H}}_{n}^{-1}\left(\boldsymbol{\beta}_{0}\right)\mathbf{X}_i^T\mathbf{A}_i^{1/2}\left(\boldsymbol{\beta}_{0}\right)\bar{\mathbf{R}}^{-1}\boldsymbol{\varepsilon}_i\left(\boldsymbol{\beta}_{0}\right) \|, i=1,\dots,n.$ We assume an ordered sequence \( h_1 \leq h_2 \leq \dots \leq h_n \), which does not restrict generality. Applying the Cauchy-Schwarz inequality,  
\[\begin{aligned}\tilde{H} &= \sum_{i=1}^{n} \left[\frac{1}{\pi_{i}} \left\| \bar{\mathbf{H}}_{n}^{-1}\left(\boldsymbol{\beta}_{0}\right)\mathbf{X}_i^T\mathbf{A}_i^{1/2}\left(\boldsymbol{\beta}_{0}\right)\bar{\mathbf{R}}^{-1}\boldsymbol{\varepsilon}_i\left(\boldsymbol{\beta}_{0}\right) \right\|^{2} \right] \\  
          &= \frac{1}{r} \sum_{j=1}^n \pi_j \sum_{i=1}^n \left(\pi_i^{-1} h_i^2 \right) \\  
          &\geq \frac{1}{r} \left(\sum_{i=1}^n h_i \right)^2.\end{aligned}\]
Equality holds if and only if \( \pi_{i} \propto h_{i} \).  
Thus, when  $\pi_i =r h_i /(\sum_{j=1}^n h_j), i=1,\dots, n,$ the condition \( \pi_{i} \leq 1 \) is satisfied, and \( \pi_i \) provides the optimal solution.

Otherwise, if \( r h_n / (\sum_{j=1}^n h_j) > 1 \), then set \( \pi_n = 1 \).  
Thus, equation \eqref{24} can be reformulated as an optimization problem for \( \pi_1, \dots, \pi_{n-1} \):  
\[\begin{aligned}
& \min\quad \tilde{H} = \sum_{i=1}^{n-1} tr \left[ \frac{1}{\pi_i} \left\| \bar{\mathbf{H}}_{n}^{-1}\left(\boldsymbol{\beta}_{0}\right)\mathbf{X}_i^T\mathbf{A}_i^{1/2}\left(\boldsymbol{\beta}_{0}\right)\bar{\mathbf{R}}^{-1}\boldsymbol{\varepsilon}_i\left(\boldsymbol{\beta}_{0}\right) \right\|^2 \right] \\  
& \operatorname{s.t.} \quad \sum_{i=1}^{n-1} \pi_i = r-1, \quad 0 \leq \pi_i \leq 1, \quad i = 1, \dots, n-1.
\end{aligned}\]  
This problem follows an iterative structure, where the optimal solution minimizes the objective function 
$\tilde{H} = \sum_{i=n-k+1}^{n} h_{i}^{2} + \left(r-k\right)^{-1} \left(\sum_{i=1}^{n-k}h_{i}\right)^{2},$ for some \( k \) such that  
\[\frac{\left(r-k+1\right)h_{n-k+1}}{\sum_{i=1}^{n-k+1} h_i} \geq 1, \quad \text{and} \quad \frac{\left(r-k\right)h_{n-k}}{\sum_{i=1}^{n-k} h_i} < 1.\]

Assume that \( T \) exists such that  
\[\max_{1 \leq i \leq n} \frac{h_i \wedge T}{\sum_{j=1}^n h_j \wedge T} = \frac{1}{r},\]  
and that \( h_{n-k} < T \leq h_{n-k+1} \).  It follows that \( \sum_{i=1}^{n-k} h_i = (r-k)T\).

Substituting \( \pi_{i}^{MV} = (\sum_{j=1}^{n} h_{j} \wedge T)^{-1} r \left(h_{i} \wedge T\right) \) into \eqref{24}, we obtain  
\[
\begin{aligned}
\tilde{H} &= \sum_{i=n-k+1}^{n} h_{i}^{2} + \frac{1}{r} \left(\sum_{i=1}^{n-k} h_{i}\right)^{2} + \frac{1}{r} \left(\sum_{i=n-k+1}^{n} T\right) \left(\sum_{i=1}^{n-k} h_{i}\right) \\  
&= \sum_{i=n-k+1}^{n} h_i^2 + \frac{1}{r} (r-k)^2 T^2 + \frac{1}{r} (r-k) k T^2 \\  
&= \sum_{i=n-k+1}^{n} h_i^2 + (r-k)T^2 = \min \tilde{H}.
\end{aligned}
\]  
Thus, \eqref{24} attains its minimum when \( \pi_i^{MV} \) is used..

Next, we show that there exists a value \( T \) such taht \( h_{n-k} < T \leq h_{n-k+1} \).  
Observe that \( k \) satisfies  
\[
\frac{\left(r-k+1\right)h_{n-k+1}}{\sum_{i=1}^{n-k+1}h_i} \geq 1 \quad \text{and} \quad \frac{\left(r-k\right)h_{n-k}}{\sum_{i=1}^{n-k}h_i} < 1.
\]  
Setting \( T = h_{n-k+1} \) gives  
\[
\frac{\left(r-k+1\right)h_{n-k+1} + \left(k-1\right)T}{\sum_{i=1}^{n-k+1}h_i + \left(k-1\right)T} \geq 1.
\]  
This leads to \((h_n \wedge T) / (\sum_{j=1}^{n} h_j \wedge T) \geq 1 / r.\) Similarly, setting \( T = h_{n-k} \) yields \( (h_n \wedge T) / (\sum_{j=1}^{n} h_j \wedge T) < 1 / r. \) Since the function \(\max_{1\leq i\leq n} (h_{i} \wedge T) / (\sum_{j=1}^{n} h_{j} \wedge T)\) is continuous in \( T \) given \( h_1, \dots, h_n \), the existence of \( T \) is guaranteed.  

On the other hand, for any \( h_n \geq T^{\prime} > T \), it follows that  \(T^{\prime} \wedge h_n \geq T \wedge h_n.\) From this, it can be derived that \( T^{\prime} / (\sum_{i=1}^n (h_i \wedge T^{\prime})) \geq T / (\sum_{i=1}^n (h_i \wedge T)). \) Thus, given \( M \in (h_1, h_n) \), the function \( (h_n \wedge T^{\prime}) / (\sum_{i=1}^n (h_i \wedge T)) \) is non-increasing. Therefore,  
\[
\max_{1\leq i\leq n} \frac{r\left(h_i\wedge T\right)}{\sum_{j=1}^n\left(h_j\wedge T\right)} = 1,
\]  
which confirms that \( h_{n-k} < T \leq h_{n-k+1} \).

Since the proof of Theorem \ref{tfour} follows similar arguments, it is omitted here.
\\
\noindent\textbf{\emph{The proof of Theorem \ref{tfive} } }
The condition \( \hat{p}_i^{sos} \geq \rho r/n \) ensures that \( \max_{1\leq i\leq n} (n\hat{p}_{i}^{sos})^{-1} = O_{P}(r^{-1}) \). The consistency of the estimator is guaranteed by Theorem \ref{theorem2}. Since \( r_1 r_2^{-1/2} \to 0 \), it suffices to focus on the subsample drawn in the second step. The primary difference between \( p_i^{os} \) and \( \hat{p}_i^{sos} \) lies in the replacement terms, namely \(\tilde{\boldsymbol{\beta}}_{r_1^*}, \tilde{\mathbf{R}}_{r_1^*}, \mathbf{H}_{r_1}(\tilde{\boldsymbol{\beta}}_{r_1^*})\), and \( \hat{\Psi} \).  
Under Assumptions (C1) to (C7), the consistency of \( \tilde{\boldsymbol{\beta}}_{r_1^*} \) follows, and \( \tilde{\mathbf{R}}_{r_1^*} \), \( \mathbf{H}_{r_1}(\tilde{\boldsymbol{\beta}}_{r_1^*}) \), and \( \hat{\Psi} \) are also consistent estimators of
\[n^{-1} \sum_{i=1}^n \left\|\mathbf{L}\mathbf{H}_{r_1}^{-1}(\tilde{\boldsymbol{\beta}}_{r_1^*})\mathbf{X}_i^T\mathbf{A}_i^{1/2}(\tilde{\boldsymbol{\beta}}_{r_1^*})\tilde{\mathbf{R}}_{r_1^*}^{-1}
\boldsymbol{\varepsilon}_i(\tilde{\boldsymbol{\beta}}_{r_1^*})\right\|.\]Thus, by Theorems \ref{theorem2} and the continuous mapping theorem, asymptotic normality is established.

\appendix

\end{document}